\documentclass{aa}  

\usepackage{graphicx}
\usepackage{txfonts}
\usepackage{natbib}
\usepackage{color}
\makeatletter
\renewcommand*\aa@pageof{, page \thepage{} of \pageref*{LastPage}}
\makeatother

\usepackage{hyperref}
\hypersetup{
    colorlinks=true,
    citecolor=blue,
    linkcolor=blue,
    filecolor=magenta,      
    urlcolor=blue,
}

\begin{document}

   \title{The UTR-2 decametre pulsar and transient survey}

   \subtitle{I. Transient detection}

   \author{V. V. Zakharenko
          \inst{1,2}
          \and
          I. P. Kravtsov\inst{1,3}\fnmsep\thanks{Corresponding author}
                    \and
          I. Y. Vasylieva\inst{1,4,5}
            \and
          P. Zarka\inst{6,7}
          \and
          O. M. Ulyanov\inst{1}
          \and
          O. O. Konovalenko\inst{1}
          \and
          A. I. Shevtsova\inst{1}
          \and
          A. O. Skoryk\inst{1}
          \and
          K. Y. Mylostna\inst{1}
            }

   \institute{
        Institute of Radio Astronomy of NAS of Ukraine, 4 Mystetstv st., 61002, Kharkiv, Ukraine\\
            \email{zakhar@rian.kharkov.ua, i.p.kravtsov@gmail.com}               
        \and
        V. N. Karazin Kharkiv National University, 4 Svobody Sq., 61022, Kharkiv, Ukraine
        \and
        LPC2E, OSUC, Univ Orl\'eans, CNRS, CNES, Observatoire de Paris, F-45071 Orl\'eans, France
        \and
        Department of Cell Biology, University of Pittsburgh, Pittsburgh, Pennsylvania, USA
        \and
        Center for Biologic Imaging, University of Pittsburgh, Pittsburgh, Pennsylvania, USA
        \and
        LIRA, Observatoire de Paris, CNRS, PSL, Sorbonne U., U. Paris Cit\'e, Meudon, France
        \and
        ORN, Observatoire de Paris, CNRS, PSL, U. Orl\'eans, Nan\c{c}ay, France
             }

   \date{Received -, 2025; accepted -, 2026}

  \abstract
    {This paper presents a detailed description of the Decametre Pulsar and Transient Survey of the Northern Sky that was carried out in 2012-2017 using the world's largest radio telescope at decametre wavelengths -- UTR-2 in Ukraine. This extensive survey covers the northern sky from declination $-10\degr$ to $+80\degr$, with a temporal resolution of 8 ms, and explores dispersion measures up to 30 pc/cm$^3$. }
    {The major advantage of the decametre wavelength range is a comparatively wide band ($\Delta f/f \sim 1$), in which the dispersive delay due to the interstellar plasma reaches hundreds of seconds, giving us the opportunity to determine the dispersion measure with a very high accuracy. This enables us to discover new transients, while avoiding data contamination from numerous weak signals of a different nature.}
    {The drift-scan survey in 5-beam mode of UTR-2 was carried out at night time. To cover the entire sky along the right ascension, the duration of the sessions was more than 12 hours at a time close to the autumn and spring equinoxes (to obtain the same conditions for the interference situation). 90 degrees along the declination were covered by five beams in $\sim 40$ days (each equinox).}
    {We discovered 380 individual transient signals with dispersion measures significantly differ from those of known sources. We determined the parameters of each single transient signal. We show that they cannot be explained by ionospheric scintillations. Repeated observations have shown that some detected transient signals are repetitive and are thus likely to originate from pulsars or rotating radio transients.}
   {}

   \keywords{Stars: neutron -- pulsars: general -- Methods: data analysis -- Methods: observational -- Astronomical databases: Surveys}
  
\maketitle
  %

\section{Introduction}

Pulsar surveys \citep{Hewish1968, Cole1968, PILKINGTON1968, Manchester1978, Manchester2001, Lazarus2015, McEwen2020, Han2021} have been carried out since the discovery of the first pulsar \citep{Hewish1968}. Recently, the number of discovered objects underwent a fast rise due to ongoing searches for regularly emitting pulsars, or other kinds of sources of pulsed radio emission \citep{Brionne2025}; for example, rotating radio transients (RRATs; \citealt{McLaughlin2006}). A series of large radio telescopes have been involved in such searches; for example, Parkes Radio Telescope, Green Bank Telescope, Effelsberg 100-m Radio Telescope \citep{Lorimer2000}, LOFAR (Low Frequency Array, \citealp{Sanidas2019}), CHIME (Canadian Hydrogen Intensity Mapping Experiment, \citealp{Amiri2021}), UTMOST (Upgrade of the Molonglo Observatory Synthesis Telescope, \citealp{Bailes2017}), and NenuFAR (New Extension in Nan\c{c}ay Upgrading LOFAR, \citealp{Zarka2015, Zarka2020}).

In \cite{Karako-Argaman2012}, the authors suppose that RRATs are a separate class of pulsars characterized by sporadic pulsed radio emission. The sporadic character of RRATs’ emission makes them difficult to detect by means of standard pulsar search algorithms, which are based on the periodicity of the signals. The researchers used the created post-processing algorithms for automatic identification of bright single astrophysical signals (such as RRATs) and this allowed them to discover 33 new RRAT candidates in the drift-scan survey data on the Green Bank radio telescope (350 MHz). Further observations of seven objects were made, six of which have already been confirmed. It should be noted that the number of RRAT candidates found is close to the number of pulsars found in that survey (34).
   
The rarely emitting nature of the RRATs studied prevents us from reliably determining their periods, and consequently whether they are indeed RRATs or transient sources of another nature. That is why this population is poorly studied. However, the growing variety of properties, such as transient intermittent scales, hints at continuum behaviour, which confirms the hypothesis of a connection between RRATs and pulsars with nulling and may be just an extreme case of the latter. The results achieved by the authors are consistent with the proposed relationship between RRATs, transient pulsars, and classical pulsars as sources in different parts of the pulse activity spectrum.

In 2010--2011 the first decametre pulsar census was carried out using the UTR-2 radio telescope \citep{Zakharenko2013}. This enabled the discovery of decametre radiation from nearly 30 pulsars, compared to a dozen of pulsars detected in the range before 2010; moreover, the census showed that a pulsar signal broadening (both intrinsic and scattering), which on high frequencies reaches $f^{-0.25}$ \citep{Cordes1978}, continues to 10--20 MHz. And even in comparison with the rather low frequency 100 MHz there is a quite significant pulse broadening: at 25 MHz compared to 100 MHz a pulse broadens by 1.6--1.8 times. This is a significant advantage of the low-frequency range, because there are preconditions for detecting pulsars or RRATs, which are undetectable at higher frequencies due to the narrower cone of radio emission. However, observations of pulsed radio emission in the decametre wavelength range are difficult for a number of reasons, the main one being scattering in the interstellar medium because repetitive pulses are `smudged' and perceived as continuous radiation. Besides, the level of the galactic background changes as ($f^{-2.55}$) and reaches brightness temperature values of $10^{5}\,\mathrm{K}$ in the Galactic disc at 20 MHz. The significant effect of scintillation and powerful earth RFI (Radio Frequency Interference) forces one to exclude `affected' channels, which reduces the sensitivity of observations. The radiometric gain is also small, since the absolute decametre band for astronomical observations is insignificant (maximum 20 MHz from $f_L$~=~10 MHz to $f_H$~=~30~MHz).

As a result, due to the modernization of the recording equipment of UTR-2 radio telescope, a decametre census of pulsars made it possible to register radio emission from 40 out of 74 pulsars with dispersion measures ($DM$) of less than 30 pc/cm$^3$, a declination higher than $-10\degr$, and a period longer than 0.1\thinspace s. During the second census of pulsars with similar parameters at UTR-2 (2020--2021), we detected decametre radio emission from 20 more sources (out of 163 $-$ 41 = 122 searched pulsars; \citealt{Kravtsov2020a, Kravtsov2022}). The detection of individual pulses of one more pulsar is described in \citep{Vasylieva2014}.
    
The continuation of this work was the first `blind' decametre survey of the entire northern sky in order to search for pulsars and transients (hereinafter -- the Survey; \citealt{Vasylieva2013, Vasylieva2015, Kravtsov2016a, Kravtsov2016b, Kravtsov2016c, Zakharenko2017, Zakharenko2018}). The search algorithm for repetitive and single pulses has significant differences. This article presents the results of searching for single pulses (transients), which can be generated by both different types of neutron stars (pulsars, RRATs, and XDINSs -- X-ray Dim Isolated Neutron Stars, \citealt{Malofeev2007}), and other cosmic sources.

Pulsars with periods of near one second can often emit anomalously intense pulses (AIPs) in the decametre range \citep{Ulyanov2006, Ul'yanov2012}, which in the survey will be shown as single (transient) events. Similarly, giant pulses from millisecond pulsars \citep{Popov2006} can be registered as solitary or sporadically occurring, while their regular pulses will be `smeared' by scattering. Moreover, the decametre range is characterized by a strong scintillation effect that usually interferes with observations but sometimes can amplify individual pulses. These reasons ignited great interest in the search for individual pulses in the decametre wavelength range.

Nevertheless, significant radio frequency interference (RFI) at low frequencies and the presence of ionospheric scintillations makes it extremely difficult to identify astronomical objects. The selection of space signals on the background of natural and artificial interference is a separate task. Only after all possible tests for the presence of terrestrial interference were conducted and the interference was excluded from the data can we talk about signal detection. This is the main aspect of this article.

The paper is organized as follows. The second section describes the Survey parameters and the data processing pipeline. In the third section we describe our tests for interference and the results of detection. The fourth section is dedicated to the discussion of the obtained results and Sect. 5 summarizes our conclusions.
  
\section{Instruments, parameters of the Survey, and data processing}

\subsection{Selected parameters of the UTR-2 radiation pattern}

The UTR-2 radio telescope \citep{Konovalenko2016} was put into operation in the early 1970s for creating a catalogue of sources in the decametre range. The T-shaped telescope configuration (east--west and north--south arms, Fig.~\ref{fig1}) provides 30$^{\prime}$ angular resolution. 
Moreover, a multi-beam system (five beams of the north--south arm) was included in the project initially for both the creation of the possibility of primary (`rough') mapping and negotiation of the influence of refraction in the ionosphere. Each antenna has a `knife' radiation pattern, where the beam width is determined from the relation $\lambda/D$, and the characteristic size, $D$, is 1968$\times$54 m for the north--south arm and 45$\times$900 m for the east--west arm, or 21$^{\prime}$$\times$12$^{\circ}$44$^{\prime}$ and 15$^{\circ}$17$^{\prime}$$\times$46$^{\prime}$ at 25 MHz to the zenith. 
Two beams of two antennas can form a combined radiation pattern $(U1+U2)$, where $U1$ and $U2$ are signals from the east--west and north--south arms, respectively, a differential antenna pattern $(U1-U2)$, and a narrow `pencil' beam at multiplication. The latter can be carried out in accordance with the formula $(U1+U2)^2 - (U1-U2)^2 = 4U1U2$ (the sum and the difference of these signals are obtained using the sum-difference devices) with additional digital processing. 
To cover the sky in the drift-scan survey, the telescope beams were tilted in the meridian plane, and the right ascension sweep was provided by the Earth rotation. When the beam is inclined from the zenith the projection (and effective) area is proportional to cosine of the zenith angle ($z$), and consequently the expanding antenna pattern. In the Survey the zenith angle did not exceed 60$^{\circ}$ (a drop in effective area by no more than two times), so the north--south antenna beamwidth by a small size was in the interval $21^{\prime}\div 42^{\prime}$ in the mean 30$^{\prime}$. In this way, five beams occupied an angle of approximately 2.5$^{\circ}$, and total angular spacing of 100$^{\circ}$ (the declination interval from --10$^{\circ}$ to 90$^{\circ}$, UTR-2 latitude is 49$^{\circ}$38$^{\prime}$10$^{\prime \prime}$N) was split into about 40 `stripes'.

In contrast to the survey of continuous sources, for which the confusion effect is significant and a `pencil' (multiplied) beam is required, in this survey we used a summary diagram (Fig.~\ref{fig1}), since there is no confusion effect for pulsars and transient signals due to the possibility of identifying signals by $DM$. This gives a significant gain in the size of the field of view and increases the probability of detecting sporadic signals. The centres of five `knife' beams of the radio telescope shifted by the beamforming system on 23$^{\prime}$ to the north (Fig.~\ref{fig1}).

Thus, approximately 40 days of continuous recordings should cover the entire area of declination from --10$^{\circ}$ to 90$^{\circ}$ and 24 hours in the right ascension. Yet considering that the interference is minimum at night, and the influence of the ionosphere is also weakened, the survey was conducted over approximately 80 nights with 6 month intervals between observation sessions close to the equinoxes to provide approximately the same interference conditions when observing different regions of the sky, 20--30 nights during 2013--2015. Since some recordings due to equipment failures turned out to be of poor quality, additional observations were carried out during 2016--2017. This was a complete (first) survey of the northern sky (the studied delta range is from --10$^{\circ}$ to about +80$^{\circ}$), during which 380 transient signals were detected. The resulting survey map is shown in Fig.~\ref{fig2} (declination interval from --10$^{\circ}$ to 90$^{\circ}$, UTR-2 latitude is 49$^{\circ}$38$^{\prime}$10$^{\prime \prime}$N).

After that, to confirm the results, a partial (second) survey of the northern sky was conducted in the spring of 2019. It covered about 10\% of the available celestial sphere, and 65 transients were detected in about two weeks of observations.

\subsection{Receiving equipment and recording parameters}

Signals of five UTR-2 beams were recorded by highly linear broadband digital receivers \citep{Zakharenko2016} with online calculation of spectra. Receivers have two channels: channel 1 for the sum of signals and channel 2 for the difference in the signals of the east--west and north--south antenna arms. The temporal and frequency resolutions as well as the maximum value of $DM$ were chosen the same as for the previous, first pulsar census \cite{Zakharenko2013} aiming to re-detect decametre pulsar radio emission. The number of frequency channels in 16.5--33.0 MHz band was 4096 (frequency resolution 4.028 kHz). The temporal resolution, $\tau$ = 7.944 ms, was chosen according to the fact that only for two known pulsars (PSRs B0809+74 and B0950+06) is the scattering time constant in the interstellar medium (at 25 MHz) less than 8 ms. Therefore, for the overwhelming majority of sources of pulsed radio emission, the scattering time constant will be bigger than the chosen temporal resolution. The $DM$ range (0--30 pc/cm$^3$) corresponds to our galactic neighbourhood up to a distance of 1--2 kpc, depending on the direction.

One of the initial results of the Survey was the very first detection in the decametre range of the nearby pulsar J0240+62 \citep{Vasylieva2014} in a single pulse mode, which is direct evidence of the capabilities of the developed software to detect individual pulses from cosmic radio sources.
For single wideband pulses (transient signals), the sensitivity at the level of $\sigma$ is
\begin{eqnarray}
        \sigma = 2kTA_{EFF}^{-1}(\Delta f \cdot \tau )^{-0.5},\label{eq1}
\end{eqnarray}
where $k$ is a Boltzmann constant and $T$ is the temperature of Galactic background + system noise. In the decametre wavelength range, the galactic background is dominating: $2\times 10^{4}\text{--}5\times 10^{4}\,\mathrm{K}$ at 25 MHz. The effective area ($A_{EFF}$) of the radio telescope is proportional to inclination and changes from 140,000 to 70,000~m$^2$. For the average temperatures and effective area $T$ = $4\times 10^{4}\,\mathrm{K}$, $A_{EFF}$ = 100,000~m$^2$, $\Delta f$ = 16.5 MHz, and $\tau$ = 7.944 ms the sensitivity at 4$\sigma$ equals 12 Jy.

\subsection{Observations and data processing}

The RFI is primarily generated by sources such as broadcast radios, ionospheric probes, discharges of car ignition systems, radio stations, and radars. The observations were carried out at night (usually from 18:00 p.m. to 7:00 a.m. local time (EET), when the interference situation is the most conducive with overlapping of two 13-hour intervals in right ascension).

Our data processing pipeline includes several stages: (i) RFI mitigation, (ii) dedispersion and selection of candidate signals, and (iii) visual analysis with adjustment of some parameters of candidate signals (primarily $DM$) in order to obtain the maximum signal-to-noise ratio (S/N) and distinguish cosmic signals from RFI. Similar pipelines have been developed for pulsar and transient studies at other low-frequency radio telescopes, such as the NenuFAR (see \citealt{Bondonneau2021}). 

\textit{RFI mitigation}. The first stage of the data cleaning is the analysis of every frequency band with the extraction of the values above 3$\sigma$ at each iteration and the definition of the mean and standard deviation following normalization and centring. To identify connected areas of increased intensity, the SumThreshod method \citep{Offringa2012} was used with the parameters chosen for the test measurements \citep{Vasylieva2013, Vasylieva2015}. The essence of the method is to use sliding windows along the frequency and time co-ordinates. When a certain threshold is exceeded in all samples of the window a mask is applied to the sequence, lately replaced by the average threshold level. There were chosen window sizes of [2, 8, 16, 128, and 256] for the frequency bands and [1, 2, 4, 8, and 64] for the time co-ordinate. The threshold for a single window was set to $T_1$ = 10, while for $N$ pixels the threshold $T_N$ was calculated:
\begin{eqnarray}
        T_N = \frac{T_1}{a^{\log_2 N}}\label{eq2}
,\end{eqnarray}

\noindent where $a$ = 1.5 is the empirical coefficient. Also, we masked frequency channels and dynamic spectra for which the average value exceeds four standard deviations for time-summed and frequency-summed channels -- time samples. An example of the resulting mask of `bad' pixels is shown in Fig.~\ref{fig3}.

    The main interference is narrowband RFI from broadcast stations (horizontal lines in Fig.~\ref{fig3}), and broadband discharges of natural and artificial origin (lightning, discharges of car ignition systems, etc.). It is extremely important to note that at this stage almost all `spots' (intensity increases) caused by scintillation effects of continuous sources are eliminated. $T_N$ for $N$ = 256 is less than 0.4 standard deviation. All connected samples that are above this value will be excluded. Therefore, intense ionospheric scintillations turn out to be strongly suppressed in further analysis.
    
    \textit{Dispersion delay compensation}. When we search for individual pulses, it is important to note that the signal delay is frequency-dependent ($f^{-2}$). Frequency-dependent scattering can also be an important parameter for identifying space radio signals, except for the decametre waves where the intensity of galactic background radiation increases significantly and the flux density of pulsars generally decreases (pulsars are one of the main types of objects searched for). With a small S/N in only part of the band, it is very difficult to determine the scattering time constant.

    The reliably registered $\Delta t$ delay of the signal (in seconds) at a lower frequency, $f_L$, with respect to a high frequency, $f_H$ (in Hz), which are the limits of the operating frequency range, was calculated as \citep{Backer1993}
\begin{eqnarray}
        \Delta t = 10^{16}DM/2.410331 (f_L^{-2}-f_H^{-2}),\label{eq3}
\end{eqnarray}
    with a high probability indicating the cosmic origin of radiation and allowing us to define the dispersion measure with great accuracy.

    At low frequencies, we can get a large relative range (the ratio of the upper frequency to the lower). For these frequencies, the dispersion delay in the interstellar plasma reaches tens and hundreds of seconds: for $f_L$ = 10 MHz, $f_H$ = 30 MHz, and $DM$ = 15 pc/cm$^3$, $\Delta t$ equals 553.18 s.
    
\begin{figure}
                \includegraphics[width=\columnwidth]{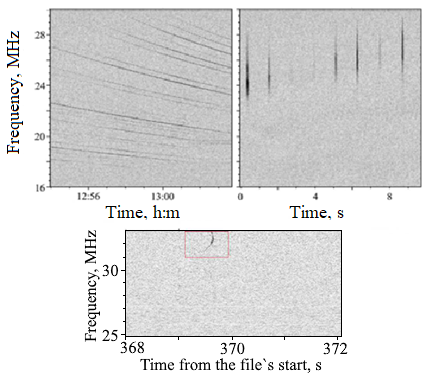}
    \caption{Top panel: Pulses of PSR B0809+74 before (left) and after dedispersion with the frequency dependence $\propto f^{-2}$ (right). On the bottom panel, a terrestrial signal with a linear frequency dependence ($\propto f^{-1}$) after dedispersion at $f^{-2}$ is shown. It radically differs from the pulses of a cosmic source after the dedispersion procedure: even in the uppermost frequency band (0.1 of the total operating range, 31.5--33 MHz), the difference in dispersion delay is tenths of a second, which is much larger than the signal width (ten milliseconds). Such interference signals are easily identified and eliminated.}
    \label{fig4}
\end{figure}

\begin{figure}
\centering
                \includegraphics[width=0.8\columnwidth]{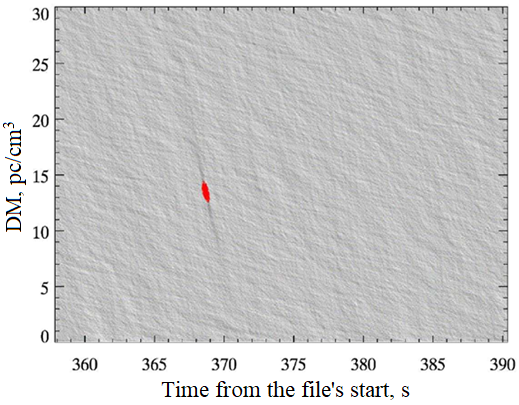}
    \caption{`Time vs $DM$' pulse plane, shown for the signal from the bottom panel in Fig.~\ref{fig4}. For the linear frequency dependence ($\propto f^{-1}$), the `dispersion measure' (when the frequency-integrated signal exceeds the set threshold 5.5 standard deviation) is close to 13.5±0.6~pc/cm$^3$.}
    \label{fig5}
\end{figure}

    A large dispersion delay at low frequencies and a high time resolution make it possible to easily isolate and eliminate the frequency-time dependence of those that differ from the law $\propto f^{-2}$. 
    With any value of the trial $DM$ after dedispersion, the pulse becomes distorted (not `vertical', the example of such interference ($\propto f^{-1}$) is shown on the bottom panel of Fig.~\ref{fig4}). 
    With a dedisperced pulse duration of several tens of milliseconds, the `non-verticality' on the panel of dynamic spectra is comparable to the pulse width; for instance, for the 5 MHz bandwidth the difference is clearly distinguishable even when the power, $n$, of the law $\propto f^{-n}$ differs from 2 only for 1\%, i.e. $n$ = $1.98\div2.02$. 
    This allows us to filter out the overwhelming number of signals that are inclined on dynamic spectra, especially considering that the target signals are broadband. The difference between quadratic and linear dependence can be also clearly seen on the `Time vs $DM$' plane of Fig.~\ref{fig5} as the widening of the detected interval of dispersion measures.

    \textit{Selection of the step for $DM$ measurement procedure.} Modelling calculations of loss of sensitivity at inaccurate coincidence of the true dispersion measure and the trial one (with which the procedure of dedispersion is conducted), under the assumption that the pulse has a rectangular shape, can be carried out by the following formula \citep{Cordes2003}:
\begin{eqnarray}
        \frac{F_{\delta DM}}{F}  = \frac{\sqrt{\pi}}{2} \zeta^{-1} \mbox{erf}(\zeta),\label{eq3p1}
\end{eqnarray}
where $F_{\delta DM}$ represents the fluxes at inexact $DM$ with the $\delta DM$ error, $F$ the flux at the true $DM$ value, and erf($\zeta$) the error function,    

\begin{eqnarray}
        \zeta = 6.91 \cdot 10^{-3} \delta DM \frac{\Delta f}{Wf} ,\label{eq3p2}
\end{eqnarray}
where $\Delta f$ is the bandwidth (16 MHz), $W$ the pulse width (100 ms), and $f$ the central frequency of observation (0.024 GHz). With these parameters and the allowable inaccuracy of the ratio (\ref{eq3p1}) not more than 5\%, the $\delta DM$ value, obtained from (\ref{eq3p1}) and (\ref{eq3p2}), should not exceed 0.005 pc/cm$^3$. This means that in the `blind' search for signals of pulsars and transients, it is desirable to have a step of the test dispersion measure no more than 0.01 pc/cm$^3$ to obtain results at insignificant losses in S/N. As $W$ decreases (which is true for a real signal), the $\delta DM$ value should decrease proportionally. Thus, for S/N maximization at the final stage of processing, the $\delta DM$ decreases up to 0.002 pc/cm$^3$.

\subsection{Search for candidate signals} 

\begin{figure}
\centering
                \includegraphics[width=0.8\columnwidth]{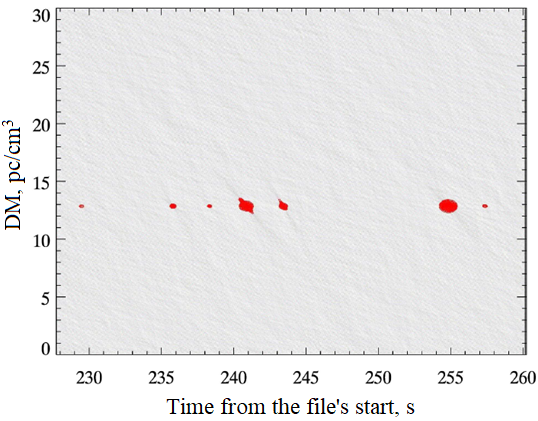}
    \caption{Result of PSR B0834+06 individual pulse detection: the sequence of red spots (each one consists of red circles) with radii proportional to S/N in their centres at $DM$ = 12.88 pc/cm$^3$. }
    \label{fig6}
\end{figure}

The next step of the pipeline is a search for signal candidates. In the `Time vs $DM$' plane, we search for the samples that exceed the threshold of 5.5 standard deviations at each $DM$ step. The samples that exceed the threshold are marked with red circles, whose radii are proportional to S/N. In Fig.~\ref{fig6}, individual pulses of the PSR B0834+06 are shown. The sequence of red spots (each one consists of red dots) for $DM$ = 12.88 pc/cm$^3$ is the result of detecting pulsar signals.

\begin{figure}
\centering
                \includegraphics[width=0.9\columnwidth]{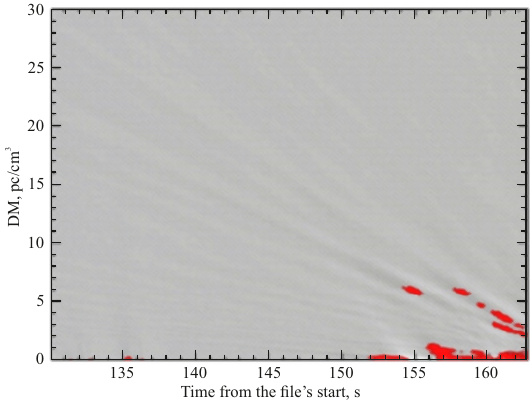}
    \caption{Example of RFI on a `Time vs $DM$' plane. The presentation of the results over the entire range of $DM$ allows us to highlight the very weak influence of noise on the $DM$ values more than 10 pc/cm$^3$.}
    \label{fig7}
\end{figure}

In Fig.~\ref{fig7}, an example of interference signals is shown. It is important to note that spectral integration allows us to significantly increase sensitivity when detecting broadband signals.

The number of images analysed on the screen, when each pixel corresponds to a certain point in time and frequency, gives several million pictures. To reduce the number of images, compression of the analysed image by 4 times in time and 3 times in the number of $DM$ steps was used. Even though the number of images was reduced to 550,000, it still required a lengthy visual analysis. 
Analysis of the effects of interference and scintillations is critical at this stage. This stage is the main one in terms of the number of detections to be found, and therefore it is vital to filter all unwanted signals. A more detailed analysis is shown in Fig.~\ref{fig7}. Integration of interference of varied duration and intensity-frequency dependence can give a very varied view when converted to the `time--dispersion measure' co-ordinates, where strings corresponding to sequential $DM$ values are not independent.

The bursts marked in red are easy to distinguish from those close to point-like ones, which individual pulsar signals give. But the actual dependence of the lines by $DM$ steps also allows us to remove from the analysis those parts of the recordings on which `rays' of increased (and/or decreased) intensity are visible, which can lead to addition with a random signal and give a false positive detection.

At this stage, all samples are eliminated by time, where even the weakest regular structures associated with interference or scintillation are visible. If the signal (that has exceeded the 5.5$\sigma$ threshold -- marked with red circles) has no signs of interference, the operator marks it as a `candidate signal'.

\subsection{Multi-parameter analysis of candidate signals} 

For the most complete and simultaneous analysis of the selected candidates for the absence of any features other than pulsar pulses and RRATs, a set of procedures for processing and analysing individual events was developed with the ability to interactively make changes to the parameters at all its stages \citep{Kravtsov2016a, Kravtsov2016b, Kravtsov2016c, Zakharenko2017}. The processing procedure displays several dependences of the signal parameters (`time--S/N', `frequency--S/N', `Time vs $DM$' plane) and provides the ability to adjust processing parameters. This allowed us to filter out a large number of interferences that were not identified by automatic processing and pre-selection.

\begin{figure}
                \includegraphics[width=1\columnwidth]{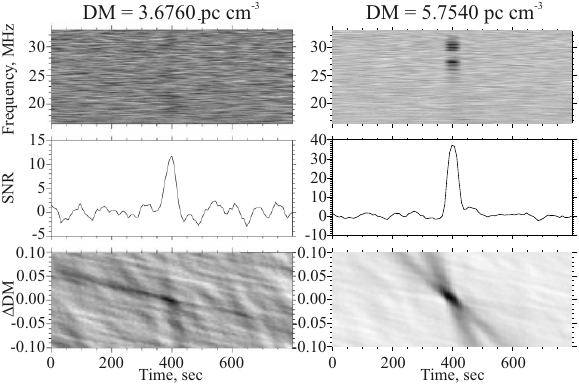}
    \caption{Example of a transient signal from an unknown source, $DM$ = 3.676~pc/cm$^3$, (left panel) and a single pulse of PSR B0809+74 (right panel), $DM$ = 5.754~pc/cm$^3$. The top panels are dynamic spectra, the middle panels are pulse frequency averaging, and the bottom panel is the `Time vs $DM$' plane.}
    \label{fig8}
\end{figure}

The most important adjustable parameter is the dispersion measure. As was mentioned above, when searching for candidate signals we used a dispersion measure step of 0.01 pc/cm$^3$. In the utility of multi-parameter analysis, the step by $DM$ was chosen to be equal to 0.002 pc/cm$^3$, which made it possible to equalize the arrival time of the pulse at all frequencies (from 33 to 16.5 MHz) with an accuracy of at least 12 ms; therefore, the dedispersion error is practically equal to the time sample (8 ms). Figure~\ref{fig8} shows an example of the detected transient signal and a pulse of a known B0809+74 pulsar.

\begin{figure}
\centering
        \includegraphics[width=0.9\columnwidth]{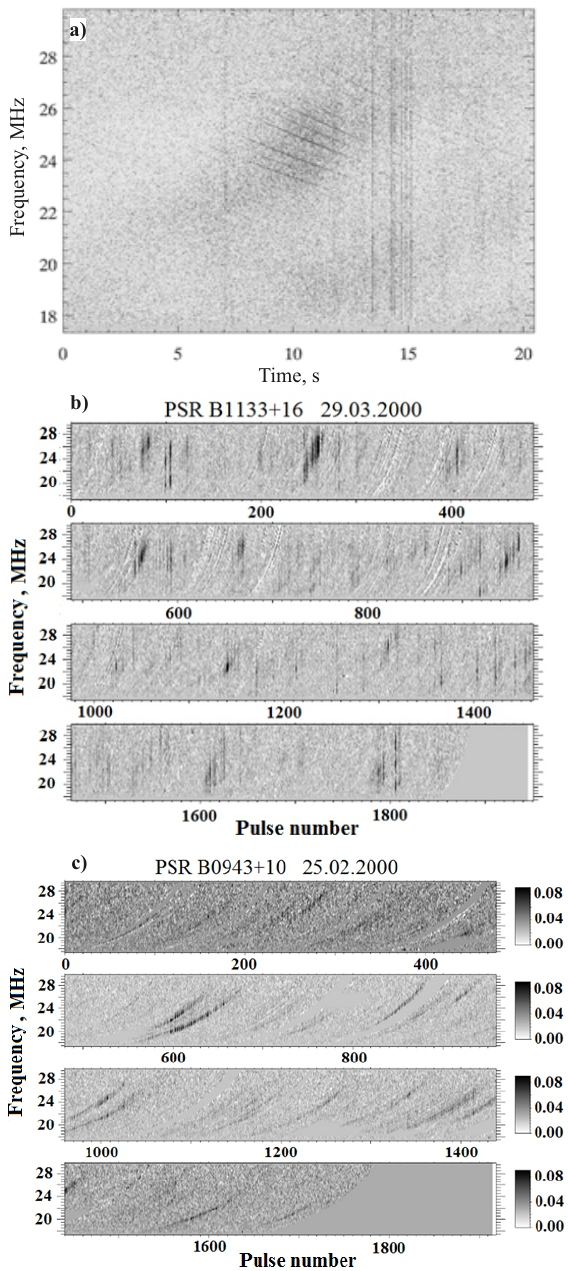}         
    \caption{Influence of scintillations upon individual pulses detection: (a) original dynamic spectra panel, in which pulsar signals (PSR B1133+16) were amplified by scintillations, (b) pulses of the pulsar B1133+16 ($DM$ = 4.84066 pc/cm$^3$) after the dedispersion procedure, in which each point corresponds to the excess of the pulse signal over the noise in the off-pulse part of the period \citep{Ulyanov2006}, (c) pulses of the pulsar B0943+10 ($DM$ = 15.31845 pc/cm$^3$) after the dedispersion procedure. It can be seen that the higher the $DM$ value, the narrower the scintillation-enhanced frequency intervals of pulsars become.}
    \label{fig9}
\end{figure}

A big advantage of the routine is the analysis of repetitive pulses. This helped us to identify and eliminate false-positive detections resulting from the superposition of weak repetitive interference (not having a cosmic nature) with random bursts. In the decametre wavelength range, due to scintillations, individual pulses of pulsars \citep{Ulyanov2006} are very rarely visible across the entire width of the band 16.5~MHz (Fig.~\ref{fig9}).

It often happens that a pulse appears in the band of a few megahertz or consists of several `fragments'. The greater the $DM$ (the greater the delay of signals at low frequencies relative to high ones), the narrower the frequency intervals of the pulsed radiation can be, for the same quasiperiod of scintillations. This can be seen when comparing the data of the detection of individual pulses of pulsar B1133+16 ($DM$ = 4.84066 pc/cm$^3$) in Fig.~\ref{fig9}b and pulsar B0943+10 ($DM$ = 15.31845 pc/cm$^3$) in the same Fig.~\ref{fig9}c. In the first case, the frequency interval of pulses `amplified' by scintillation is 3--5 MHz; in the second case, it is 1--2 MHz.

\begin{figure*}
                \includegraphics[width=2\columnwidth]{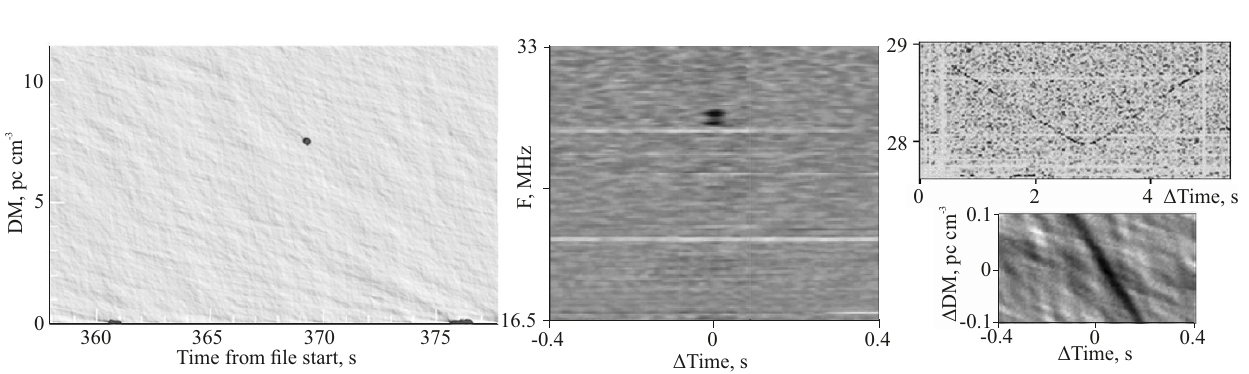}
    \caption{Example of the RFI, which is shown behind the additional multi-parametric analysis of signal candidates. In the left panel, there is a signal found by means of the previous data analysis \citep{Kravtsov2016b}.}
    \label{fig10}
\end{figure*}

\subsection{Test with `inverse' dispersion measure} 

\begin{figure}
                \includegraphics[width=\columnwidth]{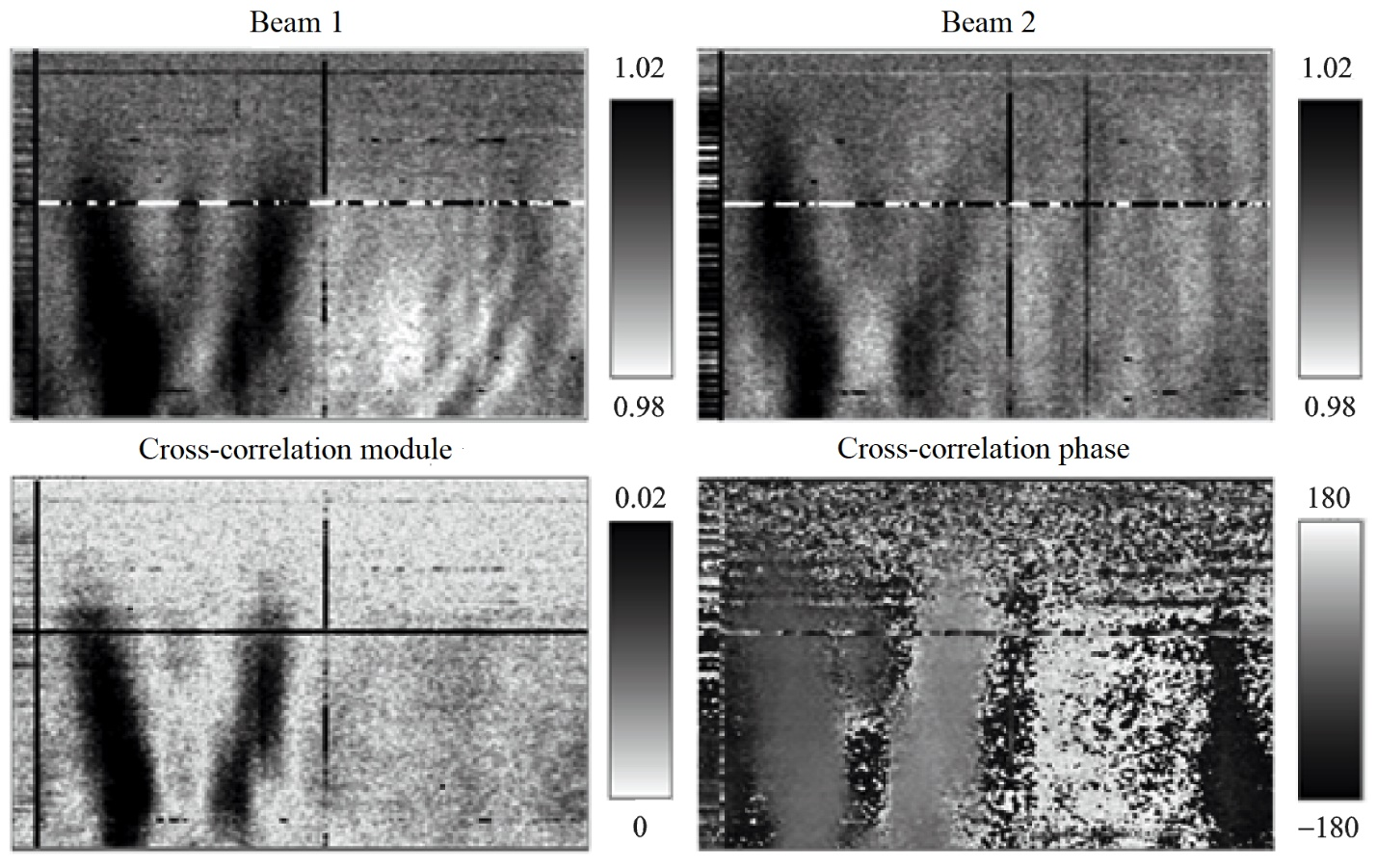}
    \caption{Examples of broadband scintillation signals showing the variety of the structures. The duration of each record is 120 s \citep{Zakharenko2018}.}
    \label{fig11}
\end{figure}

When analysing individual pulses, almost the only reliable criterion, as was noted earlier, is the dispersion delay. This eliminates a significant amount of RFI that does not comply with the law $f^{-2}$. But, some scintillations or RFI mimic signals (in the operating frequency range) with a delay in the interstellar medium. We can expect, in general, symmetrical behaviour (i.e. frequency--time dependence as $f^{-2}$ or $-f^{-2}$) of scintillations and RFI (Fig.~\ref{fig11}), since we cannot propose a mechanism for `allocating' any of the asymmetric parts of these interferences.

\begin{figure}
\centering
                \includegraphics[width=0.8\columnwidth]{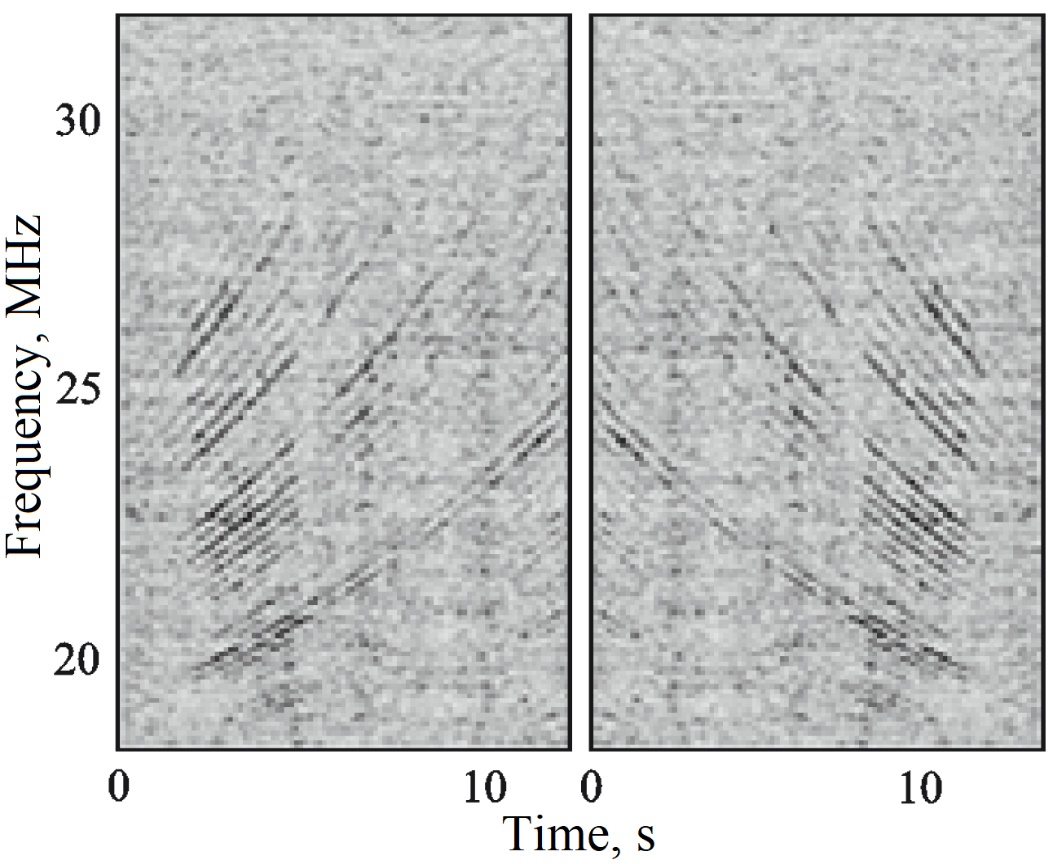}
    \caption{Inverted by the time co-ordinate signals from PSR B0809+74 for analysis with `inverse' $DM$ dependence (left). In the right panel, the initial data are given (\cite{Zakharenko2018}.)}
    \label{fig12}
\end{figure}

Despite the very thorough cleaning of the data from the effects of scintillation, it cannot be ruled out a priori that a significant number of the detections are short-time scintillations of continuous radiation of cosmic sources on ionosphere inhomogeneities (scintillations in interplanetary plasma do not result in a significant dispersion delay in the decametre wavelength range). It is known that the delay and advancement of the lower frequencies relative to the upper ones occur at scintillation with approximately the same probability. We expect the same from RFI. Therefore, a check for signals with a delay between frequencies, proportional $-f^{-2}$, may detect some signals generated by RFI or scintillations. Therefore, we proposed a test to verify the symmetry of behaviour of the detected in the Survey. Such a check will reveal the proportion of detected signals in `direct' and `inverted' recordings (Fig.~\ref{fig12}). If the part of the signals generated by scintillation or radio interference approaches 100 percent, the number of detections in both cases will be approximately equal. If most of the detected signals belong to broadband pulses with a dispersion delay, then in the `direct' recording there will be significantly more detections than in the `inverted' one.


\subsection{Criteria of natural origin of transient signals} 

Summarizing the above, we can indicate criteria that allow us to claim with a high degree of confidence that the signal is generated by a cosmic source, both for each event separately and for the entire set of detections. Sufficient broadbandness is the most significant feature of individual candidate signals, which allows us to determine the law of frequency change with time and the fact that it is close to the dependence of the delay in the interstellar medium (ISM).

The second criterion is modest or low intensity, although the signal, of course, must exceed a certain threshold. The final and valid threshold is discussed in Sect. 3.1. These requirements are not contradictory, since accumulating over a band of several megahertz, i.e. several hundreds or thousands of frequency channels of the receiver, the excess of noise level in each individual channel can be extremely small. But this criterion does not allow low-intensity interference to be distinguished from cosmic signals.

As was already mentioned, the signal broadening due to the scattering in the ISM is very difficult to detect, firstly because of the insignificant excess of S/N as in the maximum of the candidate signal, as at the level of half maximum or 1/e in the band of several megahertz (4 MHz, for example), and secondly due to the influence of scintillations (see Fig.~\ref{fig9}). Therefore, the use of this criterion is impossible as in the case of the Lorimer burst \citep{Lorimer2007}.

Further criteria are probabilistic and refer to the entire set of selected events as real space signals. Firstly, since we are dealing with small dispersion measures (less than 30 pc/cm$^3$), the sources are located in the nearest galactic environment of the Solar System, where the apparent concentration of NS (and particularly pulsars) at low galactic latitudes ($b$) is weakly manifested.

The percentage of detections in certain directions (in galactic co-ordinates, for example) will be affected by the Galactic background, whose temperature is 3--4 times higher in the direction of the disc (especially towards its centre, weaker -- to the anti-centre) than the poles. Wide beam lobes of the UTR-2 radiation pattern make this effect significant, which manifested itself, for example, in the results of detecting pulsars in \citet[see Fig. 3]{Zakharenko2013}. Thus, in the decametre range, the relative number of sources detected in the direction of the Galactic disc will be noticeably smaller than at higher frequencies. This is fundamentally different from the frequency of occurrence of interference signals, which are most common in the evening and morning, and extremely rare at night.

\begin{figure}
\centering
                \includegraphics[width=0.8\columnwidth]{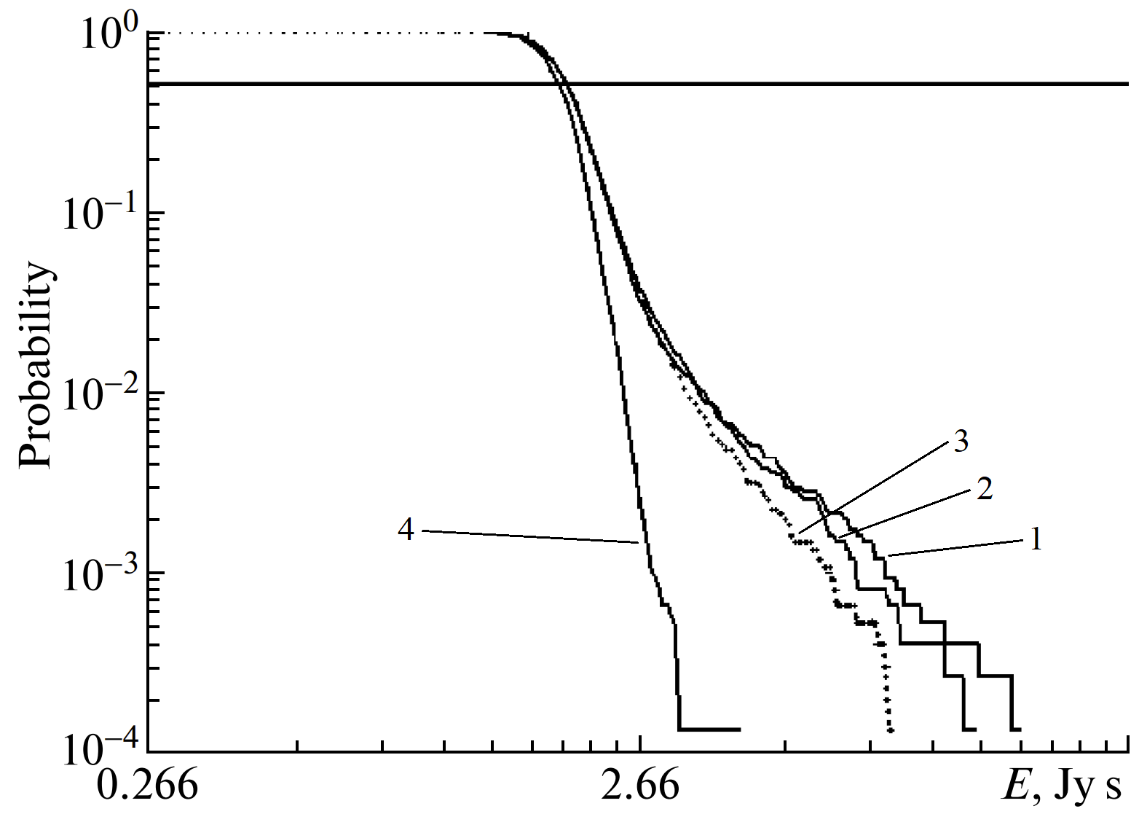}
    \caption{Energy distribution of AIPs \citep{Ul'yanov2012} of PSR B0809+74: 1,~2,~3 -- different longitudes in the average pulse profile, 4 -- out of pulse (non-pulse). In the `tail' (lines 1,~2,~3), the distribution has a character close to a power law.}
    \label{fig13}
\end{figure}

Secondly, we should expect that some of the pulses that we will be able to detect will belong to anomalously intense \citep{Ulyanov2006,Ul'yanov2012} and giant pulses of pulsars \citep{Abbate2020, Crawford2023, Mahajan2018} expanded by the influence of the ISM. Both of them have a character of intensity distribution close to a power law, shown in Fig.~\ref{fig13}. It should be taken into account that it is precisely such pulses that we can probably detect. We can also detect giant pulses of millisecond pulsars (as, for example, in \citealt{Popov2006}), and for them a power-law character is also observed in the `tails' of distributions \citep{Abbate2020, Crawford2023, Mahajan2018}.

\begin{figure}
\centering
                \includegraphics[width=0.9\columnwidth]{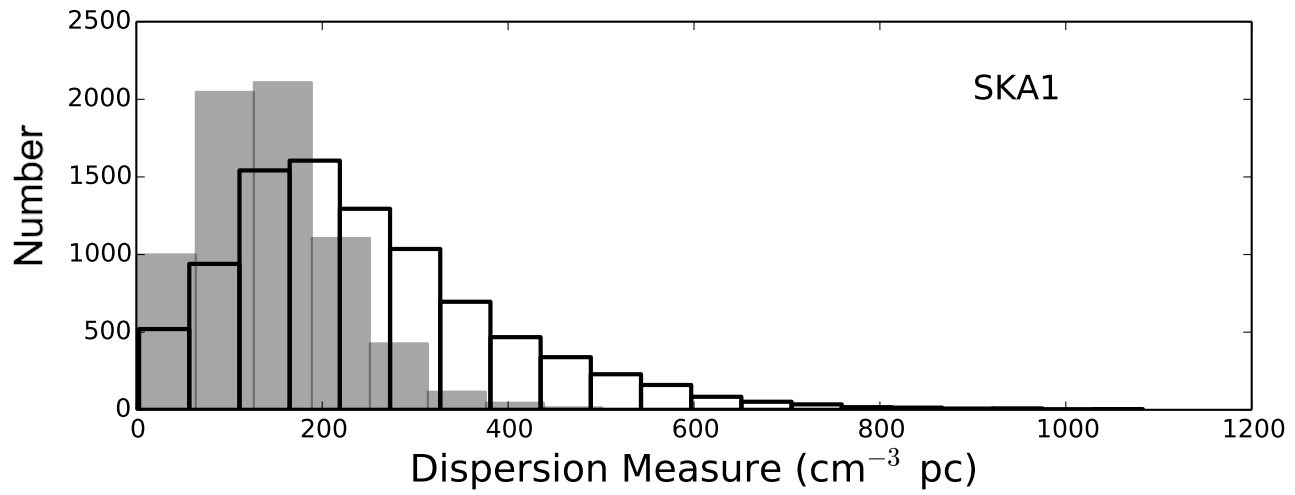}
    \caption{Histograms showing the search performance of pulsars of both SKA-LOW and SKA-MID subsystems. In SKA Phase 1, it can be seen that LOW (dark bars) performs better at low dispersion measures due to the raw sensitivity when MID (clear bars) reaches deeper into the Galaxy \citep{Keane2014}.}
    \label{fig14}
\end{figure}

Thirdly, the analysis carried out in \cite{Zakharenko2013} showed that the dispersion measure interval 0--30 pc/cm$^3$ is the most convenient for pulsar observations in the decametre wavelength range. When the $DM$ value is close to 30 pc/cm$^3$, distance-related scattering makes the pulse component hardly detectable. Moreover, the average expected signal intensity also decreases against the background of galactic noise. For small dispersion measures (small distances from the observer), a smaller number of detections is also expected since the region of space covered during observations will be insignificant. For comparison, we can use the model \cite{Keane2014} of the expected number of detected pulsars in different frequency ranges of the SKA radio telescope. To find out how to achieve the maximum possible number of discovered pulsars with the first phase of the SKA (SKA1; \citealt{Keane2014}), the authors conducted a large number of simulations that take into account all potentially available resources. Simulations included only SKA1-LOW (50--350 MHz) and SKA1-MID (350 MHz -- a few gigahertz). The $DM$ distribution of these pulsars is presented in Fig.~\ref{fig14} and it has the characteristics that were mentioned above: decreasing in the number of sources at high and low $DM$ values and a maximum at mid-values of this parameter. Given the above considerations, we can expect the maximum number of detected cosmic signals with the $DM$ values of about 10--20 pc/cm$^3$.

The fourth criterion for choosing the signals is connected with the `reverse' dispersion measure test developed by us and described above. If, in a large number of observations, the processing will result a significant excess of lagging signals compared to leading ones (low frequencies relative to the upper ones), then this excess can be suggested as the delay in the ISM, which is typical for a space-generated signals, and therefore the sought ones.

\section{Results}

As a result of processing the data of the first (full; see Sect. 2.1) decametre survey of transient signals 380 events were detected, best meeting the above selection criteria. The dispersion measure values (which can be defined with high precision and reliability in the decametre range) are unique to each of them. `Unique $DM$' means that during the Survey we excluded not only the $DM$ values of known pulsars but also all the values close to them ($\pm$0.02 pc/cm$^3$). In other words, all individual pulses from known pulsars (using the dispersion measure criterion) were removed from further analysis (as well as all detections with $DM$s close to these values, $\pm$0.02 pc/cm$^3$), so each of the 380 detections presented has its own set of co-ordinates and its own $DM$ value, which do not coincide with the parameters of known pulsars.

\subsection{Test results with the `inverse' dispersion measure and choice of detection threshold} 

\begin{figure}
                \includegraphics[width=\columnwidth]{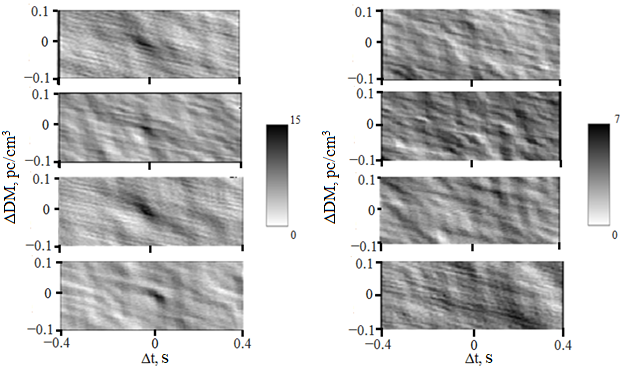}
    \caption{Two-dimensional `Time vs $DM$' dependences ($\pm$0.1 pc/cm$^3$ by $DM$ and $\pm$400 ms -- by time). In the left panel, the candidate-signals found in the Survey are shown. In the right one -- `Time vs DM' -- are dependences for the `inverse' dispersion measure (the dispersion delay is proportional to $-f^{-2}$). Candidate signals are characterized by the presence of an inclined oval of high intensity in the centre. Such a wide maximum is not observed in the case of `signals' found during the inversion of the spectrograms (adapted from \citealt{Zakharenko2018} and \citealt{Kravtsov2018}).}
    \label{fig15}
\end{figure}

For the test, 10\% of records were selected with the supposedly maximum influence of interference and scintillation with a minimum declination from --10$^{\circ}$ to 0$^{\circ}$ in the Survey. Using the pipeline 100 candidate-detections were found with a `soft' detection threshold of 5.5 standard deviations in the time-inverted data, and a multi-parameter procedure was applied to those data. The main feature was the fact of unsuccessful adjustment of the $DM$ value to increase S/N for all the detections. In the `Time vs $DM$' plane, there was no distinctive inclined oval (Fig.~\ref{fig15}). The maximum S/N for each signal in this plane was detected only in a very narrow interval, both in $DM$ and in time. Since it is typical for a random coincidence of widely spaced frequency maxima in several narrow frequency channels, we attributed such `signals' to RFI \citep{Zakharenko2018, Kravtsov2018}.

\begin{figure}
\centering
                \includegraphics[width=0.9\columnwidth]{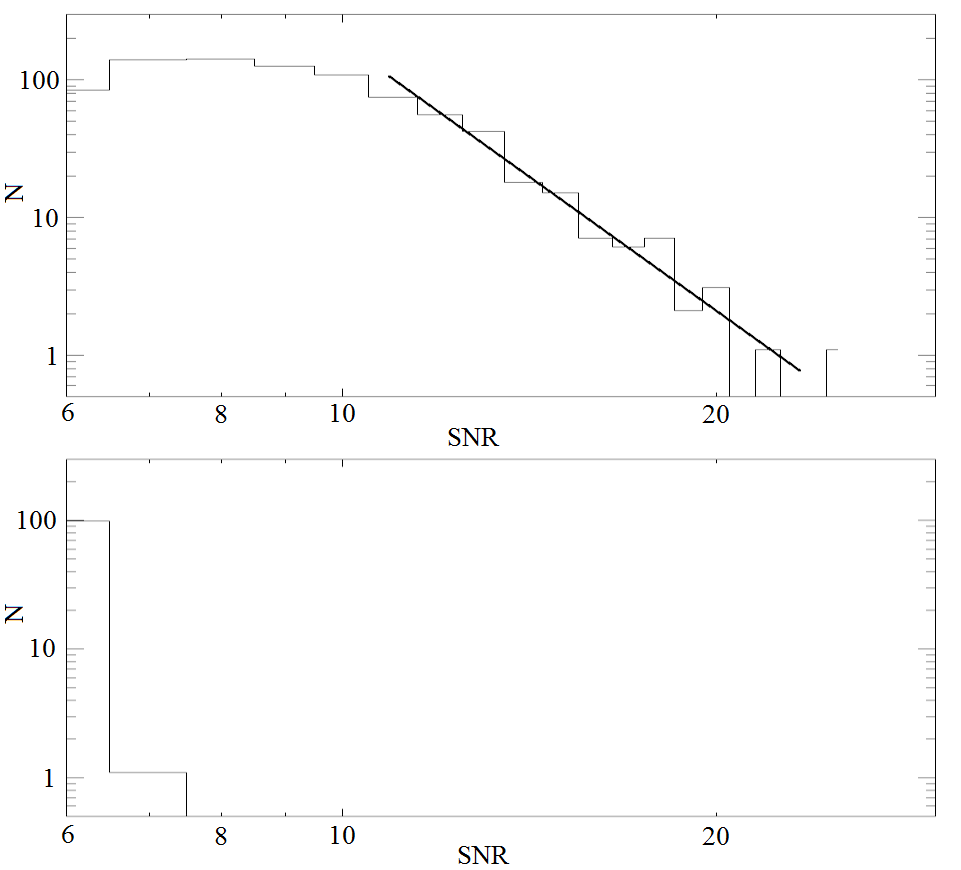}
    \caption{S/N distribution for direct and inverse dispersion measure detected candidates after $DM$-clarification. Setting the S/N threshold at the level of 8$\sigma$ makes the situation completely asymmetric: of the detected events, the only ones remaining are those for which the lower frequencies are delayed in relation to the upper ones.}
    \label{fig16}
\end{figure}

In Fig.~\ref{fig15} we separately highlight the two-dimensional `Time vs DM' dependences ($\pm$25 steps in $DM$ with a step size of 0.004 pc/cm$^3$ and $\pm$50 steps in time with an interval of 8 ms) for the signals detected with the `forward' and `inverse' dispersion measures. The figure clearly shows a dramatic difference between the candidate signals (on the left) and the signals found in the analysis with a dispersion dependence proportional to $-f^{-2}$. The characteristic oval (centre) for the candidate signals is visible in the case of each signal with a `direct' dispersion measure. For signals found with the `inverse' $DM$ ($-f^{-2}$), no such maxima are observed. Attempts to increase the S/N by varying $DM$ have practically no effect. Figure~\ref{fig16} shows a comparison of the S/N distributions for signals with $f^{-2}$ and $-f^{-2}$. Based on the test, we redefined the detection threshold, and it was finally chosen at the level of eight standard deviations.

\begin{figure}
                \includegraphics[width=\columnwidth]{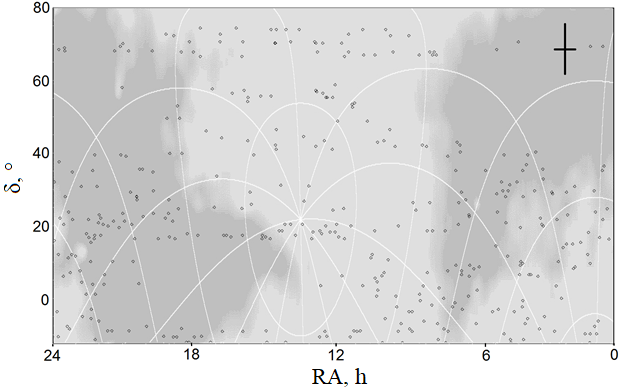}
    \caption{Sky map in equatorial co-ordinates \citep{Sidorchuk2021} showing the part of the sky targeted by the Survey and the distribution of the detected transient signals. The dots correspond to candidate signals.}
    \label{fig17}
\end{figure}

In the final stage of this method, the map was built (Fig.~\ref{fig17}); the co-ordinates of the event correspond to the co-ordinates of the centre of the cross-shaped diagram of the radio telescope. Obtained parameters, such as time from the beginning of the session, right ascension, declination, $l$, $b$, dispersion measure value, S/N, and peak flux density, are presented in the digital catalogue\footnote{\url{http://rian.kharkov.ua/images/docs/Transient_catalog.pdf/}}. The statistical parameters of the set of detections obtained from the Survey are presented in the following subsection.

\subsection{Distribution of the selected transient signals over the observational time}

\begin{figure}
\centering
                \includegraphics[width=0.9\columnwidth]{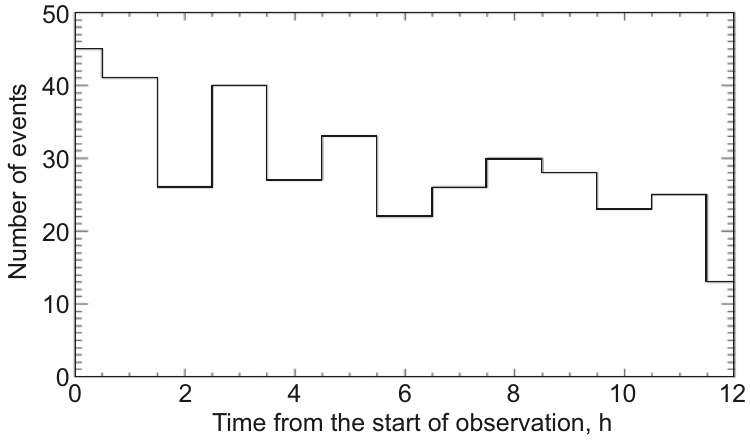}
    \caption{Distribution of the number of events depending on time of the observation session (380 detections with a S/N of more than 8). The end of the session is in the morning.}
    \label{fig18}
\end{figure}

As was previously noted, it is difficult to explain why transient cosmic signals would be detected with a roughly uniform probability throughout the observing window (18:00--07:00 EET). The galactic co-ordinate distribution is discussed below. Yet for interference, a probability distribution will have a significant excess in the evening and morning hours (with an initial threshold of 5.5$\sigma$, the number of detected RFI in these hours could exceed 100,000 per observation session). But after multi-parametric analysis, the number of candidates was reduced to a few units.
In Fig.~\ref{fig18}, the histogram of the detection distribution of the selected transient signals depending on the observation time is shown. A decline in the number of registered events in the morning hours can be explained by the fact that about 25\% of observations of the 24-hour interval at some declinations consisted not of a pair of 13 hours of observations in March and 13 hours of observations in October, but of a pair of 9--11 hours of observations in May and 15--17 hours of observations in January. This should lead to some decrease in the events detected in the morning hours, when some of the sessions were already over. Taking into account this remark and a small number of events for each hour (see Fig.~\ref{fig18}), we suppose that the obtained histogram is fairly uniform in time, which corresponds to the assumption of the cosmic origin of the selected detections. Furthermore, there is a significantly smaller number of detections that exceeded the eight-standard-deviations level in regions with high brightness temperatures, as suggested in Sect. 2.7 (see Fig.~\ref{fig17}).

\subsection{Distribution of detections by the S/N} 

The histogram of the S/N distribution (see Fig.~\ref{fig16}) in a logarithmic scale has a character very close to the power law. An increase in the S/N by 2 times leads to a decrease in the number of detected signals by 15--20 times, which corresponds to values for the power law index of 4--4.5. A similar decrease in the probability of the appearance of anomalously intense pulses (AIPs) of pulsars was obtained in the work \cite{Ul'yanov2012} for the pulsars B0809+74, B1133+16, and B0950+08, as the most typical examples of AIPs. As was mentioned above, the giant pulses of millisecond pulsars also have a power-law character in the `tails' of the distributions.
For artificial or natural interference of terrestrial origin, one could expect significantly higher intensities, and hence a wider spread in this parameter.

\subsection{DM distribution of the detections} 

The next important characteristic is the $DM$ distribution of pulses. Automatic and semi-automatic selection procedures have eliminated a huge amount of powerful radio interference. They were located, for the most part, near zero dispersion measures. The investigated $DM$ interval is in the range from 1.5 to 30 pc/cm$^3$, which corresponds to the distances from the Earth of 0.1--1.5 kpc towards the centre of the Galaxy and slightly further in all other directions.

\begin{figure}
\centering
                \includegraphics[width=0.9\columnwidth]{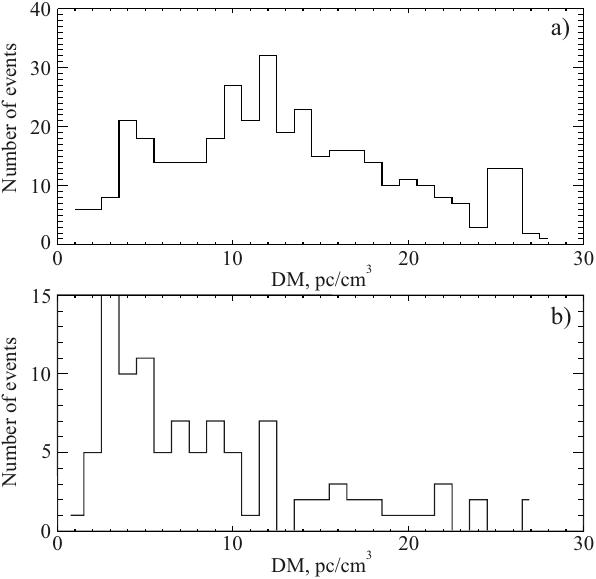}
    \caption{Dispersion measure distribution of 380 detections with a S/N of more than 8 detected in the Survey (top panel) and, for comparison, 100 `signals' with `inverse' dispersion measures and S/N in the interval $5.5\div7.11\sigma$ (it should be emphasized that all the `signals' found with the `inverse' $DM$ have a S/N of between 5.5 and 7.11 sigma, i.e. we did not remove anything from their list; bottom panel).}
    \label{fig19}
\end{figure}

Figure~\ref{fig19} shows the dispersion measure distribution for 380 selected detections (`direct' $DM$). Its character is very similar to the model calculations in \cite{Keane2014}; the maximum is about 12 pc/cm$^3$.

\begin{figure}
\centering
                \includegraphics[width=0.9\columnwidth]{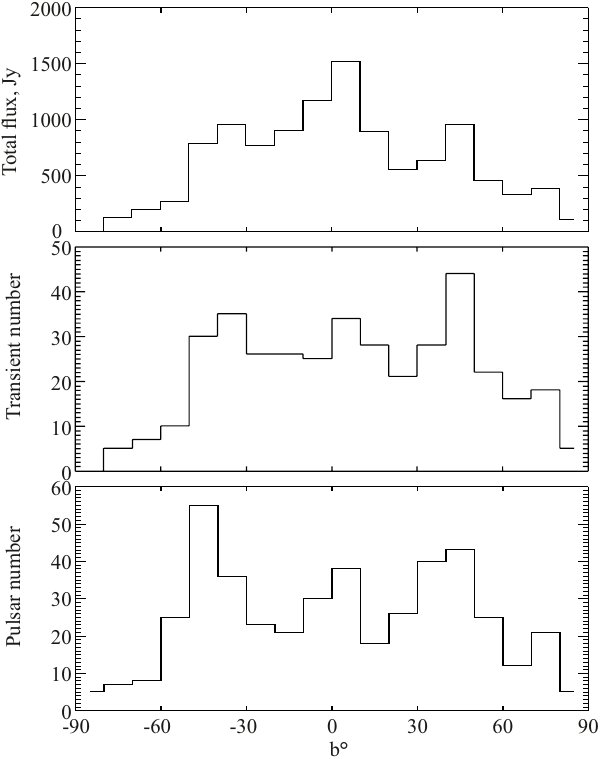}
    \caption{Galactic latitude distribution of candidate-signals, taking into account intensities of all detections (top panel). If we use only the number of events found in the Survey, the histogram does not have a peak at zero $b$ (middle panel). Close pulsars (with $DM$~<~30 pc/cm$^3$ and a period greater than 0.2~s) also have a large enough distribution by galactic latitude (lower panel, 238 close pulsars from the ATNF catalogue; \citealt{Manchester2005}).}
    \label{fig20}
\end{figure}

The lower part of Fig.~\ref{fig20} shows a similar distribution of signals detected with the `inverse' dispersion measure (none of them exceeds the value of S/N 7.11 standard deviations). It has a completely different shape. The maximum is at small values of the dispersion measure (3 pc/cm$^3$ -- 15\% of the total number of events). The same signals with $DM$ in the interval of 10--30 pc/cm$^3$ are only about 25\% of the total number. Since the maximum of the distribution tends to small dispersion measure values, its behaviour is much easier to explain by the presence of relatively broadband interference ($\Delta f\sim$ 1 MHz) with a frequency dependence close to $-f^{-2}$ lasting only a few tens of seconds.

The qualitative conformity of the form of the 380 detections distribution to the histogram in Fig.~\ref{fig14} (especially its low-frequency part) testifies in favour of the cosmic nature (see Fig.~\ref{fig19}) of the found candidates and shows considerable differences of signals with an `inverse' dispersion measure. 

\subsection{Distribution of detected signals over galactic latitude}

Another proof of the cosmic nature of the detected signals is the distribution of galactic co-ordinates. Even though neutron stars are concentrated in the central part of the disc of the Galaxy, at such small distances (hundreds of parsecs), this effect is quite weak. The distribution of close pulsars clearly confirms this (Fig.~\ref{fig20}, bottom panel).

The distribution of our transients by galactic latitude is shown in Fig.~\ref{fig20} (middle panel). The resulting distribution is quite symmetric about the disc of the Galaxy and more than half of the events are found in the latitude range of --30$^{\circ}$~<~$b$~<~30$^{\circ}$. The absence of a peak in the number of transients at zero $b$ can be explained by the fact that the detection of pulse signals in the direction of the centre and anticentre of the Galaxy at small values of $b$ is greatly complicated by a noticeable increase in the brightness of the galactic background. In part, this effect can be levelled if you use not only the number of detections, but also their intensity. The peak flux density is calculated for each event based on data on background temperatures \citep{Sidorchuk2021} and the S/N of each signal. Then (after summing the peak intensities) the histogram looks as shown in Fig.~\ref{fig20} (top panel).

\begin{figure}
\centering
                \includegraphics[width=0.9\columnwidth]{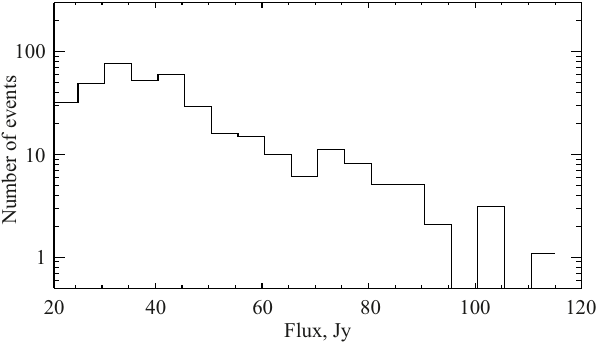}
    \caption{Distribution of maximum values of transient flux density.}
    \label{fig21}
\end{figure}

This distribution, as was expected, has a clear peak at the zero value of $b$. This also speaks in favour of the cosmic origin of the detected signals. And, since we have already obtained the data on the maximum flux density, we can analyse the intensity distribution. It is shown in Fig.~\ref{fig21} and has the same power law distribution as the S/N histogram.

\subsection{Fragment of the second survey (10\% of the sky)}

To confirm or refute the results obtained in the first `frame' of the Survey, first of all, the average frequency of transients' appearance, observations, and the entire processing cycle of part of the second `stage' of the Survey were made. After discovering a large number of transient signals in the Survey, the urgent task was to confirm such detections as reliably as possible. However, re-detection of the presence of a radiation source on a unique (single) signal is impossible. A similar problem exists for the vast majority of fast radio bursts (FRBs) observed as extragalactic sources. Nevertheless, parameter statistics -- the distribution of bursts by co-ordinates, the flux density of radio emission, and others -- can provide strong arguments for or against combining the studied phenomena into one class. For example, the homogeneity of the distribution of FRBs' co-ordinates in the celestial sphere without an emphasis on any directions in the Galaxy (such as the disc) led to the suggestion that FRBs are extragalactic phenomena.

\subsection{The parameters of the partial second survey of the sky and analysis of the data}

\begin{figure}
                \includegraphics[width=\columnwidth]{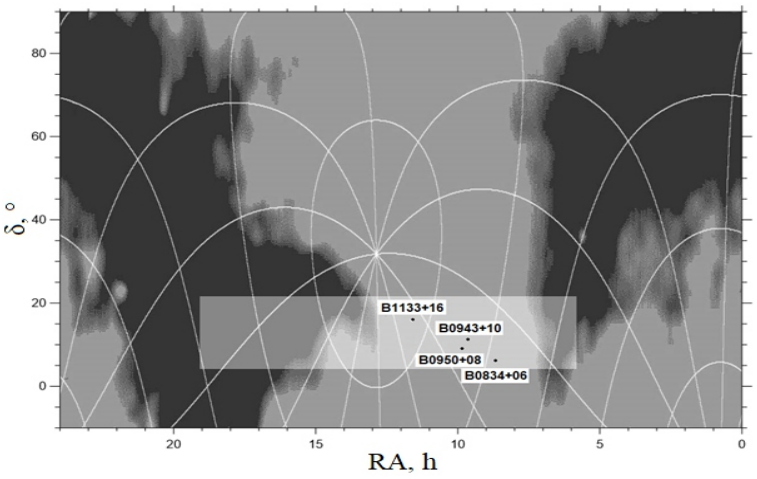}
    \caption{Sky map in equatorial co-ordinates showing the part of the sky targeted by the second partial survey (highlighted area). The dots correspond to the known pulsars -- B0834+06, B0943+10, B0950+08, and B1133+16.}
    \label{fig22}
\end{figure}

The second survey was conducted as a partial survey, based on the amount of observational data and the time required for its processing. Because the full processing of the first `stage' of the Survey took about 6 years, for the second partial survey it was selected a part of the sky, with an area of approximately 15\% of the part of the celestial sphere studied in the full survey $(\delta=-10^{\circ}\div80^{\circ})$ -- this is the area with the following co-ordinates: declination~=~+5$^{\circ}\div20^{\circ}$ and right ascension RA = 6h$\div$19h. It includes a `cold' area with several intense pulsars that can be used as calibration sources (Fig.~\ref{fig22}). The parameters of both surveys -- the total survey and the second (partial one) -- were chosen to be identical.

\begin{figure}
\centering
                \includegraphics[width=0.9\columnwidth]{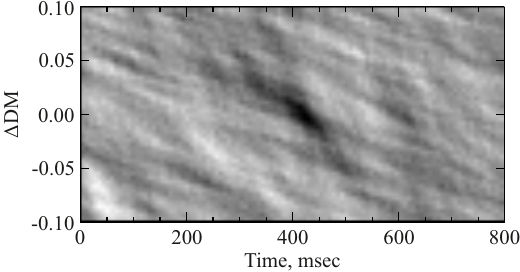}
    \caption{Transient signal V614D\_F76P5: local time of appearance 04:53:19, centre of total beam: RA = 17.3 h, $\delta$ = 13$^{\circ}$.3, dispersion measure $DM$~=~5.436~pc/cm$^3$, S/N = 10.5.}
    \label{fig23}
\end{figure}

Observations were conducted from 18 to 24 March 2019. The amount of data is about 8 TB. After completion of all stages of processing, detections with S/N exceeding 8$\sigma$ were selected. There were 65 of them.
In Fig.~\ref{fig23}, an example of the detected transient signal during the survey, which was detected in a very calm interference environment (local time of occurrence 04:53:19), is presented. It is quite broadband and stands out well on the `Time vs DM' plane.

\subsection{Comparison of the characteristics of transient signals with the results obtained in the full Survey}

\begin{figure}
\centering
                   {\includegraphics[width=0.9\columnwidth]{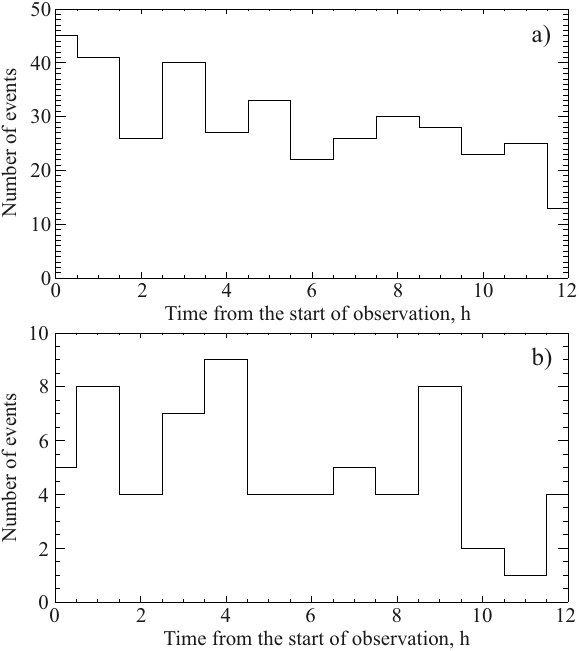}}         
         \caption{Distribution of the number of detections by the time of occurrence in the first (a, 380 detections) and repeated (b, 65 detections) survey.}
    \label{fig24}
\end{figure}

After detecting new transient signals, the distributions of their parameters, such as the local time of occurrence, S/N, and dispersion measure, were analysed and compared with the first survey. In the first survey, it was decided that the recording of the 24-hour band of the sky should be divided into two night sessions, which are separated in time for six months -- near the spring and autumn equinox. Under these conditions, the interference environment for two parts of the record (spring and autumn, 13 hours each -- for the overlapping `half-band') turns out to be quite similar. Therefore, for a further comparison, we chose the same conditions of observations -- nights near the spring equinox with the start of observations at 18:00 and the end at 7:00 of the next day. In Fig.~\ref{fig24}, the number of detections by the time of occurrence in the first (a) and repeated (b) surveys is given. It is evident that the probability of detecting a transient signal insignificantly differs from hour to hour. The difference is one and a half to two times for Fig.~\ref{fig24}a and slightly more for Fig.~\ref{fig24}b.

Therefore, it can be stated that the local time (evening and morning; which at the start of observations at 18:00 falls on the intervals 0--2 and 10--12 hours from the beginning of recording) does not cause a large number of detections that could be expected at a significant exposure to terrestrial radio interference. This is in favour of the cosmic origin of the selected detections.

\begin{figure}
                    \centering
            {\includegraphics[width=0.8\columnwidth]{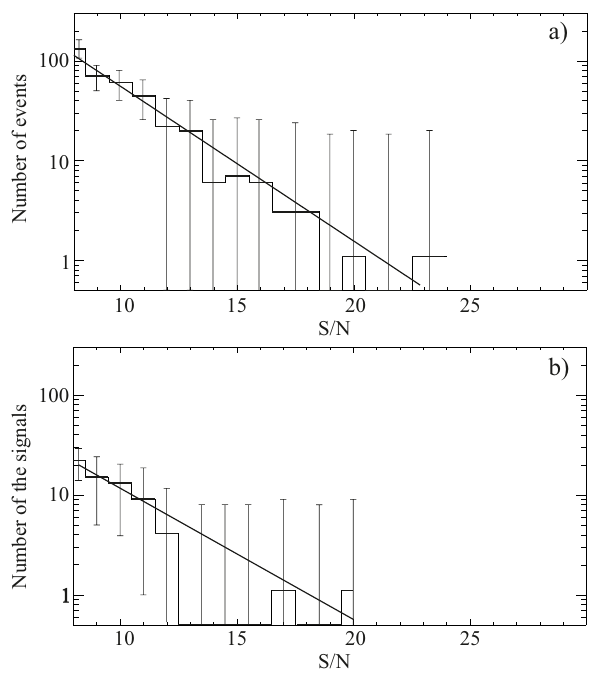}}  
            \caption{S/N for the detected signals of two surveys: the first complete (a, 380 detections) and the second partial (b, 65 detections). We believe that the `gaps' in histograms are related to the low signal statistics.}
    \label{fig25}
\end{figure}

Another characteristic is the distribution of detections by S/Ns. It should be noted that the second -- partial -- survey does not yet provide sufficient statistics. We explain the gap in the distribution of detections by S/N by the small number of sources detected in the partial survey. However, Figs.~\ref{fig25}a and \ref{fig25}b are quite similar, including the slope of the line, which is related to the value of the exponent of the power law. Therefore, we believe that the S/N distribution also indicates a high probability of the detected signals having a cosmic origin.

\begin{figure}
                  \center{\includegraphics[width=0.9\columnwidth]{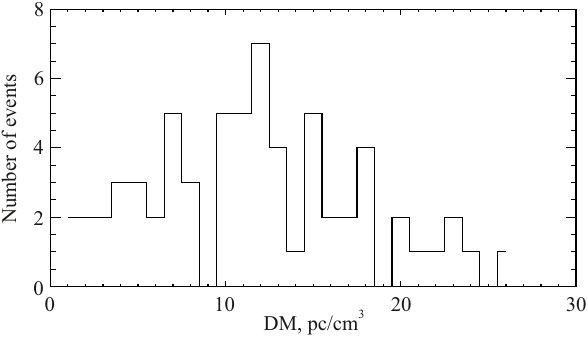}} 
            \caption{$DM$ distribution of 65 detections, found during the repeated (partial) survey. In comparison the $DM$ distribution of 380 detections with S/N > 8 detected in the Survey can be seen in Fig.~\ref{fig19}a.}
    \label{fig26}
\end{figure}

The next parameter is the $DM$ distribution of the detected signals. Figure~\ref{fig19} shows a histogram of 380 -- full Survey and Fig.~\ref{fig26} -- of the partial one (65 pulses). The shapes of the distributions are qualitatively similar and have a maximum at the same $DM$ value (12 pc/cm$^3$). The similarity between these histograms likewise convincingly indicates the cosmic origin of the detections that were made in both surveys. It is important to note that the $DM$ distribution of broadband terrestrial radio interference usually has a maximum near zero values of the dispersion measure.

It should also be noted that the burst rate of detected transient signals in the first and the second surveys was very similar: one source in about 14 hours in one UTR-2 beam. The repeated partial survey, which covers approximately 15\% of the available in survey celestial hemisphere, found 65 events, which is about 17\% of the number of events (380) that were detected in the first `stage' of the Survey. 

Thus, the characteristics of both surveys are very close. The distributions of the number of transient signals over the time of occurrence, by the S/Ns, and by $DM$ are very similar in both surveys and correspond to the expected patterns. Thus, the distribution of events by time of occurrence is close to the expected uniform distribution. At the same time, the S/N distribution is close to a power law, which is characteristic of giant pulses of millisecond pulsars or AIPs, which are characteristic of many pulsars at low frequencies. Moreover, the $DM$ distribution is also similar to that expected for both surveys. All of these results are strong arguments in favour of the assertion of the cosmic nature of all detected transient signals, or the vast majority of them.

\subsection{Accuracy of determination of co-ordinates in the Survey}

The five beams of the radio telescope allow us to make some refinements in the co-ordinates, and not only to indicate the co-ordinates of the centre of the cross-shaped directional pattern at the time of detecting the transient signal with errors of the order of $\pm7^{\circ}$ in both co-ordinates. The true co-ordinates at the time of detection could be (i) in the beam of the west--east diagram, (ii) the north--south beam, or (iii) their intersection. Since the west--east antenna beam is common in all cruciform patterns, five transients of the same time and $DM$ would be detected (see examples of such detection for PSR J0243+63 in \citealt{Vasylieva2014}). This has never been observed, which is explainable by the smaller effective area of the west--east antenna. Such detections could have been revealed at the first stage of the analysis, but in the end they did not exceed the detection threshold and did not make it into the final list. Therefore, we can say that the transient signals from the final list were received either by the north--south antenna or by the centre of the diagram, and the error in determining the co-ordinates for all detections is $\pm 15^{\prime}$ in declination (on average) and about 6$^{\circ}$30$^{\prime}$ -- in right ascension. Some of the transient signals (probably the most intense ones) could have hit the centre of the `cross', which can be checked using a difference diagram (there will be no signal). Then the error in determining the co-ordinates would be about $\pm 15^{\prime}$ in both co-ordinates. But for such a large number of detections, this check is of independent interest, and it is carried out as part of the program for re-detecting the found transient signals.

\section{Discussion}

Such a large number of single-pulse detections with unique dispersion measure values (which reliably differ from each other and the DM values of known pulsars) warrants further analysis. It is very probable that in addition to the truly transient signals, single intense pulses from the following sources were found in the Survey:

   \begin{itemize}
      \item New pulsars or RRATs;
      \item Known and recently discovered pulsars or RRATs with incorrectly defined $DM$ values that prevent them from being unambiguously compared with the data obtained in the Survey;
      \item `Unfavourably' oriented (for high frequencies) pulsars;
      \item X-ray dim isolated neutron stars (XDINSs);
      \item Anomalously intense pulses or giant pulses of MSPs;
      \item Pulsars with a steep spectral index (for example, --3), and therefore with a reduction in the flux density at high frequencies (e.g. PSR B0943+10 from 80 to 600 MHz);
      \item Some unknown sources, both of cosmic and terrestrial origin;
      \item All of the aforementioned objects, unfavourably oriented for high frequencies.       
   \end{itemize}

One of the results of the re-detection program launched at UTR-2 for the most intense of the detected transient signals in the Survey was the repeated detection of an unknown source with a dispersion measure value of $DM$~=~7.02~pc/cm$^3$. Detection during one of the repeated observations \citep{Kravtsov2020b} of pulses with $DM$~=~7.03~pc/cm$^3$ and $DM$~=~6.97 pc/cm$^3$ (the error of determining the $DM$ value of a few hundred parsecs per cubic centimeter for a signal with a low S/N is quite admissible) indicates the probable first detection of a new pulsar or RRAT in the decametre wavelength range.

Rapid growth in the number of known pulsars with $DM$ values of less than 30~pc/cm$^3$, a declination > --10$^{\circ}$, and a period of more than 0.1~s (74 sources in 2010, 163 -- in 2020; \citealt{Kravtsov2020a, Kravtsov2020c, Kravtsov2022}; due to the surveys \citealt{Good2021, Sanidas2019, Kawash2018, Dong2023}) shows that a significant number of the detected signals may belong to this population of the Galaxy. As the radiation cone of pulsars expands approximately by 1.6--1.8 times \citep{Zakharenko2013} even in the case of close frequencies (100 and 25~MHz), the number of pulsars available for observation in the decametre wavelength range increases (depending on the angle between the magnetic axis and the axis of rotation, the growth rate can range from linear to quadratic). To verify this (i.e. compare with the signal parameters detected in this Survey), we need to know the $DM$ of recently discovered pulsars with an accuracy of better than 0.01~pc/cm$^3$.

An even greater number of supernova remnants \citep{Keane2008} are RRATs. Assuming that their number is 2--3 times greater than that of pulsars, we can talk about several hundred such sources potentially available for decametre observations. As of now, slightly more than one hundred RRATs have been detected at high frequencies\footnote{\url{https://rratalog.github.io/rratalog/}}, but they have never been observed in the decametre range. The latter statement applies only to the individual pulses of these sources \citep{Karako-Argaman2015}, since during the second decametre census of pulsars on the UTR-2 radio telescope \citep{Kravtsov2022} a J2325--0530 source was detected as a pulsar, when accumulated with a period of $P$~=~0.868735115026~s and a dispersion measure of 14.958(20)~pc/cm$^3$. In addition, since the beginning of 2022, Ukrainian low-frequency radio telescopes have been searching for both individual and accumulated pulses from the RRAT, and our next papers will be devoted to these studies.

Another type of object is an XDINS. The number of these is also considered significant (almost the same as the number of `normal' radio pulsars). It is possible that XDINSs also emit single pulses in the low-frequency wavelength range \citep{Malofeev2007}, and a number of them were detected in this survey.

There are also a number of pulsars that have a steep (about --3) spectral index and, accordingly, a decrease in flux density at high frequencies. Such sources may also be unavailable for high-frequency observations, while contributing to an increase in the number of detections (possibly by their AIPs) in the decametre range.

It is also necessary to include millisecond pulsars with giant pulses in the applicants for detection. Due to the scattering in the ISM, short giant pulses expand sufficiently for their detection, and the absence of GP repeating with the same frequency as `normal' pulses (tens and hundreds of hertz) makes it easier to distinguish the pulsed component from galactic noise.

\begin{figure*}
                \includegraphics[width=2\columnwidth]{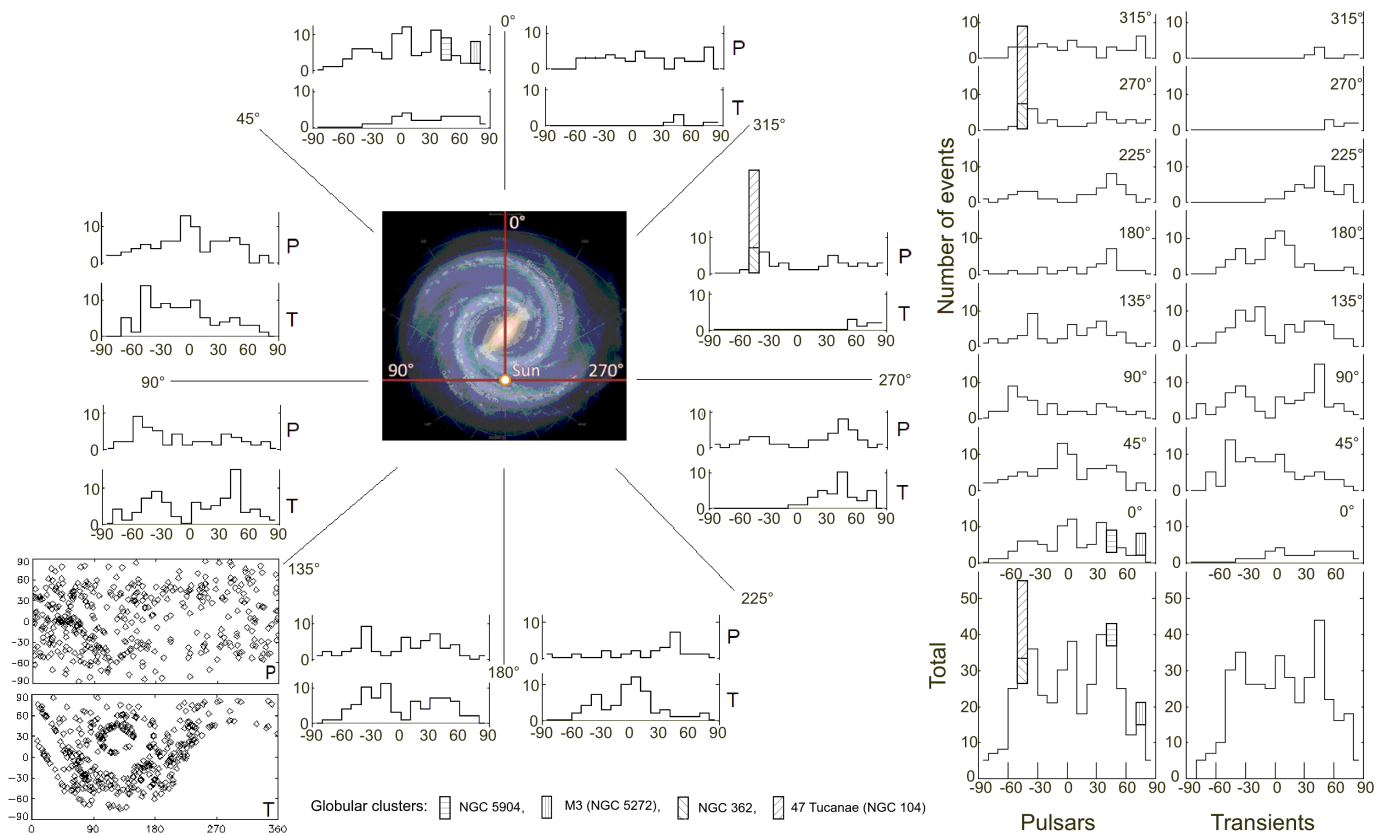}
    \caption{Galactic latitude distributions of pulsars with $DM$~<~30~pc/cm$^3$ \citep{Manchester2005} and the (probably) cosmic sources found in this work as a function of galactic longitude. The lower left panel shows galactic co-ordinates of the pulsars and transients. On the right one, the total histograms of the $b$ distributions are presented. The number of pulsars in the globular clusters NGC 5904, Messier 3 (NGC 5272), NGC 362, and 47 Tucanae (NGC 104) is indicated by different types of hatching.}
    \label{fig27}
\end{figure*}

We also need to pay attention once again to the distribution of the transients and pulsars by galactic latitude. First, thanks to many new surveys, the number of pulsars has more than doubled over the decade. The new surveys largely cover all galactic latitudes, $b$ (not just those close to the Galactic disc), which is important when comparing the $b$ distributions with the locations of the (presumably) cosmic sources we detected. Figure~\ref{fig27} shows that the distributions are similar in characteristic features: 
\begin{itemize}
    \item fairly uniform between galactic latitudes --50$^{\circ}\div50^{\circ}$,
    \item small peaks at latitudes --50$^{\circ}\div-30^{\circ}$, --10$^{\circ}\div10^{\circ}$, 30$^{\circ}\div50^{\circ}$.
\end{itemize}

These characteristic features can be observed at all galactic longitudes. Since the distribution of pulsars (due to filling the entire plane of galactic co-ordinates) is more revealing, it can be noted that there is some asymmetry in the distributions at longitudes 90$^{\circ}\div180^{\circ}$ and 180$^{\circ}\div270^{\circ}$: in the first case, sources with $b$ in the range --50$^{\circ}\div$--30$^{\circ}$ predominate, while in the second case the galactic latitude has a positive sign with approximately the same (30$^{\circ}\div50^{\circ}$) values. Whether these are signs of a particular galactic structure could be investigated with sufficient statistical data, which is a strong argument for continuing to search for new pulsed sources in surveys with increasing sensitivity. It should also be taken into account that the nature of the distributions is affected (as mentioned in Sect. 3.5) by the fact that both the number of pulsars and the background temperature increase in the disc of the Galaxy (compared to high latitudes), which interferes with their detection. This is true for any low-frequency radio telescope. In addition, the electron density at the selected $DM$ limit value constrains the study area (and the number of neutron stars in it) in the disc to about 1.5 kpc, while for high $b$ -- for example, pulsars in globular clusters (approximately ~10\% of the total) -- pulsars are located at much greater distances: from 4.5~kpc of 47 Tucanae (NGC 104) to 10.2~kpc of M3 (NGC 5272; see Fig.~\ref{fig27}). The scattering is also higher at low $b$, which also leads to a decrease in the number of detected sources at low frequencies.

Also, transient signals of various natures (which are different from the radiation of neutron stars) should not be excluded from consideration: from observations of the radio frequency `echo' of NS merging or powerful lightning on exoplanets, similar to the ones detected in the atmosphere of Saturn \citep{Zakharenko2012}, as well as other sources of unknown nature of both terrestrial and cosmic origin. Thus, one can expect the detection of signals (transient or repetitive) from hundreds or thousands of sources of various natures, with the dispersion measure value up to 30~pc/cm$^3$.

Of course, in the absence of the possibility of repeated observation of a transient signal, a natural means for increasing the reliability of research is a high effective area of the radio telescope. As the results of the intensity distribution of the transient signals detected in our work show, a 2-fold decrease in the effective area (equivalent to an increase in the detection threshold to 16~RMS) would lead to a 20-fold decrease in the number of detected events.

The presence of two or more large radio telescopes, spaced apart, which are not subjects to the same local interference and ionospheric conditions, will give much greater detection reliability. This is how a search is performed using two LWA arrays \citep{Stovall2015, Taylor2016}. However, for now, the sensitivity of each of the LWA array \citep{Taylor2012, Ellingson2013} antennas may not be enough to detect pulses similar to those found in our Survey.

Therefore, to solve this problem, radio telescopes with the largest effective area (the search is planned to be carried out jointly at UTR-2 and NenuFAR; \citealt{Zarka2015,Zarka2020}) and with a very broad band are preferred. If there is a competition over the choice of either broadband or field of view (beamlets in LOFAR and NenuFAR), then, in our opinion, broadband should be preferred. This increases the sensitivity in a cube or the fourth power -- the probability of detecting transients, based on our data on the distribution of detections by S/N.

Large radio telescopes (such as SKA) will be created in the future. With a planned field of view 100 times larger and a sensitivity 15 times higher than Arecibo, SKA will revolutionize the study of transients at radio waves. But since SKA will only be operational by $\sim 2030$, the search for single pulses is important for the development of algorithms for separating cosmic signals from a huge number of interferences of artificial and natural origin. With more sensitive tools and more sophisticated analysis, the coming decades will undoubtedly provide a much better understanding of the nature of flaming objects in radio, including known types of objects and those that will be discovered.

\section{Conclusions}

   \begin{enumerate}
      \item According to the results of the first decametre survey of the northern sky to search for pulsars and transients, 380 single wideband transient signals with unique dispersion measure values, which meet the known criteria of their cosmic origin, and which have S/Ns of more than 8, were found. The high accuracy of the dispersion measure determining allows us to exclude pulses of known pulsars from the set of detections we made, so that the detected signals were definitely generated by other sources.
      \item As a result of the proposed and conducted `inverse' dispersion measure test, it was convincingly proved that the detected signals cannot be explained by the effect of ionospheric scintillations of continuous emission of cosmic sources.
      \item The distributions of the detected signals' parameters by the dispersion measure, galactic latitude, time of appearance, and intensity of the detections are completely different from the typical artificial and natural terrestrial RFI characteristics, and this fact indicates that the vast majority of detected transient signals have a cosmic origin.
      \item A second partial survey covering 15\% of the northern sky has also revealed 65 transients, giving approximately the same average detection rate of similar detections -- one in 14 hours in one beam in the first and second partial `stage' of the Survey.
      \item Re-detection of the signal (J0337+3937) indicates the possibility of the first discovery of the source of repeated pulsed radiation -- a pulsar or RRAT -- in the decametre wavelength range, which was initiated by research conducted in this work. The obtained results also initiated the start of the second `stage' of the Survey of pulse signals of cosmic origin and the second census of pulsars for the possible identification of the transient signals detected in the Survey.
      \item Undoubtedly, the most effective method of searching for transients is two- and multi-antenna search. The use of the developed programs and methods of analysis and the use of highly sensitive radio telescopes such as LOFAR and NenuFAR in collaboration with UTR-2\footnote{UTR-2 is currently partly destroyed by Russian occupiers in 2022. Nevertheless, NAS of Ukraine and the Institute of Radio Astronomy of NAS of Ukraine are working both to preserve the surviving parts of the UTR-2 and GURT radio telescopes and to begin rebuilding them.} (and in the future -- with SKA) will substantially improve the reliability of detection and accuracy of determining the co-ordinates of sporadic radiation sources. In the future, using these instruments, the maximum number of transient sources will be able to be found by repeatedly searching for new detections. The developed set of means, such as search algorithms, methods of observations, and processing obtained data, leads to an increase in the effectiveness and informativeness of the research, and applying it to the task of investigating the nearest stellar environment is invaluable.
   \end{enumerate}

\begin{acknowledgements}
      Part of this work was supported by the Project: 101131928 -- ACME -- HORIZON-INFRA-2023-SERV-01 and by Ukrainian Program `Scientific and scientific and technical (experimental) work in the priority area `Radiophysical and optical systems for strengthening the defence capability of the state' for 2025-2026 -- `Global monitoring of radio signals of natural and artificial origin of decametre-metre waves in the interests of cosmology and applied problems of defence capability' (state registration number -- 0125U000879) as well as by means of Project 6435 of The Science And Technology Centre In Ukraine (STCU) `Peculiarities and interaction of the events in the solar corona and interplanetary medium' (2025-2026, project code `Corona'). IPK acknowledges the support of Collège de France by means of `PAUSE -- Solidarit\'e Ukraine' programme and of NAS of Ukraine by a Grant for Research of Young Scientists of the NAS of Ukraine (2019-2020, project code `Neytron' and 2025-2026, state registration number - 0125U002903, project code “Dyspersiia”), as well as the State Education Development Agency of Republic of Latvia through the Latvian State Fellowship for Research (in 2020 and 2021-2022). IPK would like to thank the Paris Observatory and the CNRS (LPC2E) for being great host organizations for the PAUSE program. IPK and AIS also acknowledge the support of the NAS of Ukraine through the Grants of the NAS of Ukraine for Research Laboratories and Groups of Young Scientists of the NAS of Ukraine (2020-2021 -- project code `Pulsar' and 2022-2023 -- project code `Spalakh').
\end{acknowledgements}

\bibliography{library-trans}
\bibliographystyle{aa}

\begin{appendix} 
\onecolumn
\section{Additional figures}

\begin{figure*}[ht!]
    \centering
    \includegraphics[height=0.43\textheight,keepaspectratio]{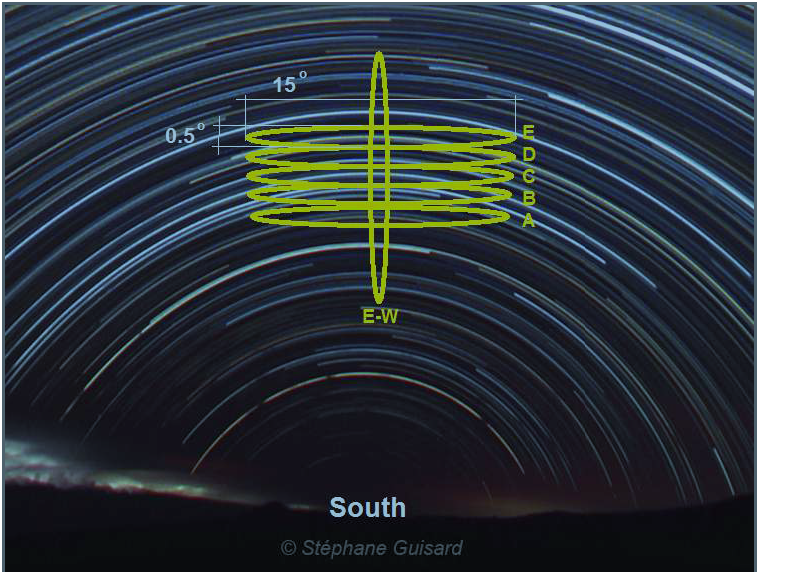}
    \caption{Five beams of the north--south antenna (A, B, C, D, E) and the beam of the west--east antenna in the direction of the celestial meridian.}
    \label{fig1}
\end{figure*}

\begin{figure*}[ht!]
    \centering
    \includegraphics[height=0.43\textheight,keepaspectratio]{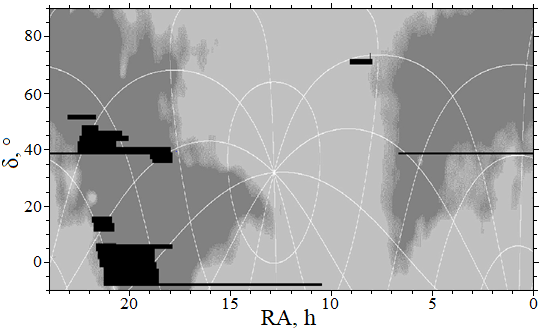}
    \caption{Sky map of the Survey with areas indicated in black that were not covered. Grayscale map of the Galactic background radio emission at 20 MHz \citep{Sidorchuk2021}.}
    \label{fig2}
\end{figure*}

\begin{figure*}[ht!]

                \includegraphics[width=\columnwidth]{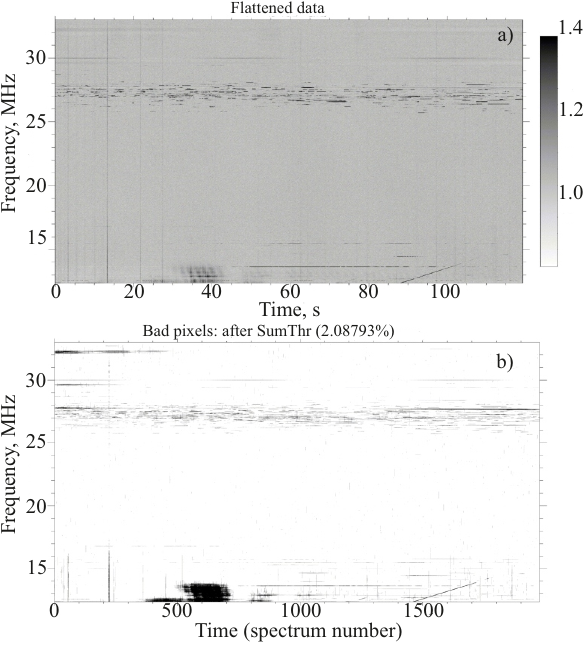}
       \caption{Normalized (flattened) dynamic spectrum (upper panel) and the resulting bad pixels mask after RFI flagging by SumThreshold: $\sim$2\% of the pixels have been flagged \citep{Vasylieva2015}.}
    \label{fig3}
\end{figure*}

\end{appendix}

\end{document}